\documentclass[preprint,12pt]{aastex}

\shorttitle{Signatures of Outflows}
\shortauthors{Tobin et al.}

\newcommand{\rsun}{\mbox{$R_{\sun}$}}

\begin{document}

\title{Imaging Scattered Light from the Youngest Protostars in L1448:
Signatures of Outflows}
\author{John J. Tobin\altaffilmark{1,2}, Leslie W. Looney\altaffilmark{1}, Lee G. Mundy\altaffilmark{3}, Woojin Kwon\altaffilmark{1}, Murad Hamidouche\altaffilmark{1}}
\altaffiltext{1}{Department of Astronomy, University of Illinois, Urbana, IL 61801}
\altaffiltext{2}{Current Address: Department of Astronomy, University of Michigan, Ann Arbor, MI 48109; jjtobin@umich.edu}

\altaffiltext{3}{Department of Astronomy, University of Maryland, College Park, MD 20742}
\begin{abstract}

We present deep IRAC images that highlight the scattered light emission around many of the youngest protostars, the so-called Class 0 sources, in L1448.  By comparison of the data with a Monte Carlo radiative transfer code \citep{whitney2003a}, we demonstrate for the first time that the observed infrared light from these objects is consistent with scattered light from the central protostar. The scattered light escapes out the cavity, carved by molecular outflows, in the circumstellar envelope. In particular, we observe prominent scattered light nebulae associated with the Class 0 sources: L1448-mm, L1448 IRS 2, and 3B, as well as a Class I source: IRS 3A. We use a grid of models with probable protostellar properties to generate model spectral energy distributions (SEDs) and images for bands
sensitive to this scattered light: J, H, Ks, and \textit{Spitzer} IRAC bands. By simultaneously fitting SEDs and images of the outflow cavities, we are able to model geometric parameters, i.e. inclination angle and opening angle, and loosely constrain physical parameters. The opening angle may be an important
indicator of the evolutionary state of a source. We compare our results for Class 0 sources to similar studies of Class I sources. There may be a transition phase from Class 0 to Class I when a source has an opening angle between 20$^o$ to 30$^o$. It is important to note that while the best fit model image and SED do not fully describe the sources, the fits generally describe the circumstellar structure of Class 0 sources in L1448.
\end{abstract}


\section{Introduction}
The L1448 dark cloud is located on the western side of the Perseus molecular cloud complex at a distance of $\sim$250 pc \citep[e.g.][]{enoch2006}. The relatively isolated, young stellar population and their associated spectacular molecular outflows make L1448 an ideal star formation laboratory. L1448 contains several embedded YSOs at the Class 0 \citep{ward1993} and Class I stage, identified as such from infrared to millimeter observations. In this paper, we study four sources in particular: L1448-mm \citep{curiel1990, bachiller1990, bachiller1991, bachiller1995}, L1448 IRS 2 \citep{olinger1999,wolf2000, olinger2006}, and L1448 IRS 3A and 3B \citep{curiel1990,looney2000,looney2003}. All are Class 0 protostars except IRS 3A, which is a Class I protostar and binary companion to IRS 3B \citep{curiel1990, terebey1997, looney2000, looney2003}. In the literature these sources have been identified with differing nomenclatures: L1448-mm is also known as L1448C; L1448 IRS3 is also known as L1448N, and IRS 3A and 3B are also called N(A) and N(B), respectively.

Outflows associated with star formation are a ubiquitous phenomenon. Observations and theory demonstrate that the outflow originates within 100 AU of the central source \citep[e.g.][]{shu1995}. With outflows originating deep within the circumstellar envelope, it is certain that the outflow will affect the evolution of the YSO \citep{arce2006}. While the launching mechanism is still uncertain, the result of an outflow emanating from deep within the circumstellar envelope is the creation of a cavity in the envelope. As the protostar ages, the outflow cavity seems to widen. The widening is perhaps initially caused by precession of the outflow jet; however the precession angles are found to be much narrower than the observed opening angle \citep[e.g.][]{reipurth2000, arce2004}. The stellar wind must be the dominant mechanism for creating wide angle cavities \citep{arce2004}. As cavities are widened throughout the Class 0 and Class I stages of star formation, the widening of the outflow cavities may act to dissipate the envelope and halt mass infall \citep{shu1987, padgett1999, arce2004}. 

We examine these outflow cavities because, in general, Class 0 objects cannot be directly detected short ward of 10$\mu$m. The cavities allow light from the embedded central source to escape and scatter off dust in the cavity at near-infrared (NIR) wavelengths. NIR emission allows us to look into the cavity to the inner portions of the envelope. This bipolar cavity and outflow structure has been studied for more than 25 years \citep[e.g][]{snell1980,kaifu1987, mundt1987}; however, recent advances in instrumentation have enabled us to explore this structure in detail for Class 0 sources \citep[e.g.][]{noriega2004}. In the past, only Class I sources could be studied in detail; high extinction in the circumstellar envelope greatly attenuates light at J, H, and Ks-bands for Class 0 sources. The Infrared Array Camera (IRAC) \citep{fazio2004} on the \textit{Spitzer Space Telescope} enables us to see through the dust enshrouding a Class 0 source. For the first time, we are able to observe scattered light from Class 0 sources with detail in multiple bandpasses. IRAC has channels numbered 1 through 4, corresponding to central wavelengths of 3.6$\mu$m, 4.5$\mu$m, 5.8$\mu$m, and 8.0$\mu$m wavelengths respectively.

Radiative transfer codes \citep[e.g.][]{whitney1992, lucas1997, whitney2003a} can be used to model scattered light in the cavity from the embedded protostar \citep[e.g.][]{lucas1997, padgett1999, eisner2005, terebey2006, stark2006}, at fluxes within the sensitivity of IRAC. Ground-based telescopes can detect scattered light emission in the J, H, and Ks bands with sufficient depth of field, requiring very long integration times. However, ground-based studies have mainly focused on \textit{only} scattered light from more evolved, Class I, sources \citep[e.g.][]{whitney1992, whitney1993, lucas1997, eisner2005, stark2006}, Class 0 sources have not been studied in detail. The cavities as observed by IRAC will detect scattered light as well as thermal and molecular emission (if present), e.g. polycyclic aromatic hydrocarbons (PAHs), CO, and H$_{2}$. Scattered light from the Class 0 sources in L1448 is primarily only present in Ks and IRAC bands; J and H bands are very marginally detected. 

In this paper, we demonstrate that the idea of scattered light cavities is valid for Class 0 sources in addition to Class I sources. Also, for the first time modeling of Class 0 sources is performed primarily using the scattered light SED and images. Modeling these sources with the radiative transfer code by \citet{whitney2003a} enables us to determine geometric parameters (inclination and opening angle) of the outflow cavity and loosely constrain physical parameters. To determine the true opening angle, we must simultaneously fit inclination. The parameters as constrained by the model are then compared to outflow observations at millimeter wavelengths. Finally, we discuss the idea of opening angle being a possible indicator of evolutionary state.

\section{Observations and data reduction}
\subsection{IRAC and near-IR observations}
We observed the L1448 cloud with two separate pointings of IRAC on 2005 February 25, one to cover L1448 IRS 3 and L1448-mm, and another to cover L1448 IRS 2. IRS 3 was observed in the High Dynamic Range mode with frame times of 30 seconds in 1 x 2 mapping mode and a cycled dither pattern of 31 positions and small scale factor achieving a total integration time of 930 seconds. The IRS 2 field was observed with the same parameters but with 30 dither pattern cycle positions for a total integration time of 900 seconds. The difference in total integration times between the two fields is not significant.

The L1448 fields had already been covered by the \textit{cores2disks} (c2d) Cycle 1 Legacy program \citep{evans2003}, but our science goals required much greater depth of imaging to be sensitive to scattered light in the outflow cavities. The additional depth has the added effect of a mitigating cosmic ray hits, producing a very clean final image with high signal to noise (S/N) ratio. This is necessary for the study of the outflow cavities, whose extended emission in these fields can be confused by the presence of background emission from the molecular cloud itself, as well as the HH objects associated with multiple YSOs in the region.

Post-BCD pipeline products were solely used in this paper. Initially it was assumed that further reduction would be required, but after receiving the data, it was decided that the quality of IRAC data from the post-BCD pipeline was very high and sufficient for our science goals. Our data were processed by pipeline version S13.2.0. A color-composite image of the combined L1448 IRS 3 and IRS 2 fields is shown in Figure 1. Figure 2 shows a grayscale image of IRAC channel 2 with the positions of the major sources plotted. Close up views of the individual sources plotted with the PAs of CO outflows and source positions are shown in Figures 3, 4, and 5. 

Additionally, the paper makes use of data from the COordinated Molecular Probe Line Extinction Thermal Emission (COMPLETE) survey \citep{complete2006}. The NIR fields of L1448 were taken in J, H, and Ks bands, from the Calar Alto Observatory 3.5m telescope with the Omega2000 NIR prime-focus imager. The observations and data reduction are described in \citet{foster2006}, a color-composite of the data are shown in Figure 1 of \citet{foster2006}. We calibrated the J, H, and Ks maps of L1448 to 2MASS standards \citep{2mass2006}. The error in flux calibration for J was 15$\%$, H 11$\%$, and 12$\%$ for Ks. The inaccuracy of the photometric calibration may partly be due to the effect of cloudshine \citep{foster2006}.

Photometric measurements of the four sources were corrected for extinction using the reddening law for infrared wavelengths \citep{rieke1985, indebe2005}. The extinction toward the sources comes from extinction maps of Perseus produced by the COMPLETE survey. The maps were made using 2MASS data and have 5$^{\prime}$ resolution. The extinction toward each source and their positions are listed in Table 1.

\subsection{IRAC and near-IR photometry}
We performed aperture photometry on each source listed in Table 1 using the Image Reduction and Analysis Facility (IRAF). Photometry of the sources was not entirely straightforward since a background annulus could not be used in this situation due to extended emission surrounding each source. Instead, using the imstat procedure in IRAF, we measured a large area of sky adjacent the source that was devoid of stellar emission due to high extinction to obtain an average background value per pixel. This value was then multiplied by the aperture area measured by IRAF and subtracted from the photometric value of our source.

For each source, we measured a 1000 AU aperture radius centered on the positions listed in Table 1. Perseus is at a distance of roughly 250 pc, and the IRAC pixel scale is $1\farcs2/\rm{pixel}$, so we measured an aperture radius of 3.33 pixels. The pixel scale of the NIR images is $0\farcs45/\rm{pixel}$, making our aperture 8.89 pixels. We limited photometry to 1000 AU apertures to minimize contamination of the SED from extended H$_{2}$ emission, HH objects, and binary companions. Photometric values are listed in Table 2.


\section{Modeling}
To interpret our data, we used the radiative transfer code of
\citet{whitney2003a}. This is a multiple scatter and radiative
equilibrium code using the Monte Carlo method. The model assumes a
rotationally-flattened envelope \citep{terebey1984} with power law
density r$^{-3/2}$ \citep[cf.][]{looney2003}. We set a small centrifugal radius of 3 AU and an
inner envelope radius of 3 times the dust destruction radius. A small
centrifugal radius assumes little rotation and near spherical symmetry,
and the dust destruction radius varies depending on particular central
source parameters. The envelope also includes bipolar cavities; the
cavity can be conical (streamline cavities) or they can be curved. We
model curved cavities as this shape best fit our observations. Also, the
model assumes a flared circumstellar disk, we use a disk with a radius of
50 AU; the inner radius is equal to the dust destruction radius. Embedded
within the disk and envelope is the central source, defined by a stellar
atmosphere and a given mass, radius, and temperature. The central source
parameters may be somewhat unphysical for Class 0 sources as the central source
is still a protostar and not a pre-main sequence star. Observational constraints
on the central source properties are not available for Class 0 sources; therefore, we use the  constraints that are available for Class I sources. \citet{nisini2005} measures effective temperatures of a few Class I sources to range from 3580 to 4900K and stellar radii of 1.2 to 2.9 \rsun. These observed Class I central sources are similar to the central sources modeled in our grid. We assume a constant effective temperature of 4000K and vary the stellar radius.

With each run, the standard mode of the program outputs SEDs at 10 inclinations with bins centered at cos \textit{i} = 0.05, 0.15, ..., 0.95, with bin widths of 0.10. The code also has a "peeling-off" option \citep{yusef1984,wood1999}, which outputs a SED at \textit{only} the specified inclination, without averaging over bins, and a high S/N image. The "peeling-off" option is much more computationally intensive than the standard mode, which is why we do not simply run the entire initial grid with this option.

In modeling our data, we first compared our photometric measurements to SEDs from a grid of models run in the standard mode. Because of the large number of free variables, we only varied: luminosity, envelope mass, envelope radius, inclination angle, and opening angle, listed in Table 3. Opening angle refers to the half opening angle of the source; the angle as measured from the central axis to the cavity edge. To scale the stellar luminosity of the sources, we ran models with eight stellar radii (R$_{*}$) at constant stellar temperature. Simply scaling a lower stellar luminosity model to higher luminosity does not give viable results because the inner disk and envelope radii depend on R$_{*}$; scaling the luminosity would imply a higher temperature of the central source. However, accretion luminosity is inversely proportional to R$_{*}$; we keep a constant accretion rate in our model grid rather than maintaining a constant accretion luminosity.

This first grid was run in standard mode with only 1 million photons per run. This reduces computation time as there are 120 independent models per luminosity. Each model takes approximately 10 minutes to run on a 2.4 GHz Opteron processor with g77 compiled code. Once finished, we convolved the SED with filter functions for common bandpasses. We used a $\chi^{2}$ routine to obtain fits of the model SED to the observed SED at the IRAC and Ks bands taken with 1000 AU apertures; the sources are too weak at J and H bands. We do not include longer wavelengths because published photometric measurements are almost always upper limits due to their apertures encompassing multiple sources (our sources are all close binaries). Also, we are not attempting to fit disk properties with our grid as the envelope is most important in regards to scattered light. The disk properties are more important at longer wavelengths e.g. submillimeter and millimeter, due to dust grain growth in the disk \citep{wolf2000, Dalessio2001, whitney2003a}. In Figures 6, 7, and 8, we show the SED of the overall best fit model with our photometric measurements, we include photometric values from the literature in the SED for completeness.

We define a good fit as being within the 90$\%$ confidence level. Our maximum allowed value is $\chi_r^{2} =$ 2, if we only fit IRAC bands, and 1.85 if Ks-band is included. We fit IRAC and Ks bands for L1448-mm and IRS 3B. IRS 2 was fit using only IRAC bands because it could not be fit at the 90$\%$ confidence level when Ks-band was included. Because IRS 2 is the weakest source at Ks-band, it is possible that there is higher extinction towards IRS 2; this would affect Ks-band photometry more than IRAC photometry. IRS 3A was unable to be fit at the 90$\%$ confidence level with IRAC or Ks bands. These fits at the 90$\%$ confidence level are then taken to be inputs for detailed modeling. At the 90$\%$ confidence level, this first phase of modeling reduced our 9600 possible models to 36 models (see Table 4) for detailed modeling. The SED fitting does not tightly constrain physical parameters, however, we do observe certain trends from SED fitting.

One trend that is observed from the SED fits at Ks and the IRAC bands is that luminosity is not well constrained by our observed NIR SEDs. A model with the same physical parameters can appear with 3 or more luminosities. This occurs because Class 0 YSOs emit the most flux in mid-infrared to submillimeter wavelengths. The luminosity is loosely constrained because fitting at only the shorter wavelengths is analogous to fitting a dog by its tail: you only roughly know if the dog is a great dane or chihuahua. Also, variations of opening and inclination angle can allow the SED to fit multiple luminosities. For example, a model that has a lower luminosity than the observed source can be a possible fit if the model has a wider opening angle and/or a smaller inclination angle.

Another trend is the recurrence of inclination angle in multiple fits. For some sources, the same inclination angle is fit for several different luminosities, envelope radii, and masses. This reaffirms that the luminosity is not well constrained by our SEDs but also that the envelope radii and masses are not well constrained either. These trends are the reasons that this first phase of modeling only gives us a baseline for more detailed modeling. In order to confirm that a model fits the data, we must also compare a model image to the observed source. When the images are compared, many of the SEDs fit at the 90$\%$ confidence level are eliminated because images produced with the same parameters do not match the data.

Our second phase of modeling used the ``peeling-off'' option of the Whitney code run with the parameter range of the SEDs fit at the 90$\%$ confidence level, producing a high S/N ratio image and an exact SED for a given inclination. The ``peeling-off'' option requires many more photons to produce a high S/N image. We ran our models with 10 million photons. Each model took approximately 6-8 hours on a 2.4 GHz Opteron processor. The images help us constrain the inclination angle and opening angle of the source. Different inclinations change the amount of flux measured from the source; flux increases from edge-on to pole-on viewing, and the appearance of the scattered light in the cavity changes \citep{whitney2003b}.

\section{Discussion}
In modeling our sources, we had good success fitting SEDs and then refining these parameters with image comparison. However when we attempted to fit these images quantitatively to the data, we had mixed results. In general, the sources have complex morphologies when compared to our ideal models. Figure 9 displays images of the models which best match our data. The general trend is that the models overestimate flux in J, H, and Ks-bands, roughly match the flux in IRAC bands 1 and 2, and underestimate the flux in IRAC bands 3 and 4. There are also morphological details of the data that cannot be fit by our models. 

Our models represent the most ideal scenarios of star formation, e.g. isolation, spherical symmetry, no external influences, etc. The data tell a different story. The sources we model have many complex morphological details not accounted for in an ideal scenario. All the sources are binary or multiple systems and the circumstellar envelopes of the sources are not necessarily spherical, the may be prolate or oblate \citep[e.g][]{bonnell1996}. In addition, molecular emission at NIR wavelengths is nearly always associated with YSOs in the form of HH objects, and emission from the disk or envelope around the protostar itself. Our models do not include flux from molecular lines present in NIR bands, PAHs, CO, and H$_2$ in particular. H$_2$ lines are present in all IRAC bands decreasing in intensity going toward longer wavelengths, the CO line is present in IRAC channel 2. On the other hand, PAH emission is not likely to be strong because the low mass YSOs in L1448 do not emit the far-ultraviolet flux necessary to excite PAH emission \citep{allaman1989, peeters2004}. Despite the possibility of line emission, spectra from the Infrared Spectrometer (IRS) \citep{houck2004}, onboard the \textit{Spitzer Space Telescope}, do not show evidence of line emission from any of the sources (IRS Disk Team, private communication). Unfortunately, IRS does not cover IRAC channel 2, thus we do not know if a CO line is present in any of these sources.
 
Figures 1 to 5 show a large number of HH objects associated with outflows in close proximity to the sources. The presence of outflows and HH objects from several sources in a small area leads to the possibility of outflows influencing the circumstellar environment of a source. Because our sources do not exist in the ideal environment of the models, we are not able to quantitatively compare their images. Nevertheless, despite the many ways in which the sources do not fit an ideal scenario, the models do generally describe the circumstellar structure of the protostars in L1448, and a qualitative image analysis can be performed.

Our primary method of matching model images to the data is the comparison of flux versus position angle at a given distance from the central source. We use an annulus of radius 3000 AU (10 pixels) from the source and a thickness of 600 AU (2 pixels). The flux density is measured at 64 discrete locations, or segments, along the annulus by a custom FORTRAN program. Only full pixels within a segment are used to measure the average flux density per segment, not partial pixels, so there is some flux lost in this measurement process. However, the loss is not crucial since the same loss is in both the models and the data, and the plots are only used for qualitative comparisons rather than numerical fitting. 

For comparison purposes, we rotate all of our cavity images such that their outflow position axis is vertical, 1.57 radians (90$^o$). Starting from a position of zero radians at the horizontal, the top of the image is positive angle and the bottom is negative angle. The center of the cavities are located at approximately 1.57 radians  for the blue-shifted side and -1.57 radians (-90$^o$) for the red-shifted side. Depending on the presence of confusing sources, we may not have used the full angular range for fitting. From the plots shown in Figures 10 to 12, it is clear that our model flux values do not entirely match the data flux values. 

The most important features we examined for model and data agreement were: how well the boundaries of the cavity match in the models and data and how well the amount of flux agreed in the IRAC bands for these plots. The boundaries of the cavity are where the flux starts to continually increase from the zero level up to the peak and then drop back to the zero level. A model with the correct combination of opening angle and inclination will match the boundaries in the data. A larger or smaller opening and/or inclination angle will not match the data and not be considered a fit. There is some degeneracy between the inclination and opening angles, however, we use the SED fitting to disentangle them. 

We then looked at how well the flux from the cavity at its peak levels matched the models. The general pattern of the flux in the models is to rise to a peak level, remain approximately at this level for the extent of the cavity, and then decrease back to zero. This pattern is observed with IRS 3B, but IRS 2 and L1448-mm rise to a peak in the center of the cavity, then drop back, covering the same angular extent as the model. The models do not simulate this feature. It may be evidence of more scattering at the center of the cavities of IRS 2 and L1448-mm than at the cavity-envelope interface, this is discussed in more detail below. For L1448-mm and IRS 2, we looked at how well the flux at the peak overlapped the maximum flux level of the model. This method does not quantitatively constrain the physical parameters, but the SED does quantitatively fit the physical parameters. Matching the images is simply a further constraint on the SED fit. It is fair to say that there could be more possible SED and image fits within the resolution of the model grid; we adopt the resolution of the model grid as our error bars for inclination and opening angle. Table 4 describes the possible uncertainty of fit parameters in more detail.

As mentioned previously and shown in Figures 10 to 12, there are inconsistencies between the model and data that are dependent upon wavelength. A possible reason for these inconsistencies is that there may be more extinction toward the sources than derived from published extinction maps. We base this assertion on the fact the models over produce flux at the J, H, and Ks-bands. We tried different variations of A$_{V}$ attempting to match the J, H, and Ks-bands. The appearance of the models began to better agree with the data if we apply an A$_{V} =$ 20. However as we better match the NIR bands, our models and data diverge in the IRAC bands. Also, at high extinctions, our observed SEDs do not fit any models. An alternative explanation for these inconsistencies could be incorrectly assumed dust properties.

Incorrectly assumed dust properties seem more likely than an extinction of A$_{V} > 20$. \citet{whitney2003a} uses interstellar medium (ISM) dust models in the cavity and envelope. An ISM dust model may not be completely realistic for protostellar envelopes. The dust albedo of the ISM models used by \citet{whitney2003a} increases steadily at wavelengths short ward of 10$\mu$m and peaks around 1$\mu$m. If this and the published values of A$_V$ are correct, there should be much more scattered light than we observe in the J and H bands, and less scattered light in IRAC channels 3 and 4. We do not observe this trend. Also, we do not know the size of the dust grains in the observed cavities or the effects of dust grains varying in size. Larger dust grains could cause more scattering in IRAC channels 3 and 4 and less scattering in the J and H bands. Additionally, cavity density is assumed as a constant in our models and not varied in the grid, however we did explore the effects of half an order of magnitude and an order of magnitude higher or lower cavity density than listed in Table 3. There is a very important caveat to modeling a constant cavity density, partly depending on the opening angle. A cavity density an order of magnitude greater than our assumed value begins to have a mass that is not negligible as compared to the mass of the envelope.

A greater cavity density than our assumed value (see Table 3) decreases scattered light at short wavelengths (J, H, Ks, and IRAC channels 1 and 2). A lower cavity density than our assumed value does not increase or decrease scattered light at short wavelengths. IRAC channels 3 and 4 are not affected by higher or lower cavity density; the model predicts that nearly all emission in these two channels is thermal due to the ISM dust model used for the cavity and envelope. From examining the effects of varying cavity density, the models are predicting that most scattered flux should come from the envelope-cavity interface rather than dust grains in the cavity; the dust grains in the cavity seem to extinct flux more than they scatter it. As discussed previously, the model does not necessarily reflect the morphology of scattered light in our data. Our SED fitting at a 1000 AU aperture is not greatly affected by higher or lower cavity density. However, an SED measured with a larger aperture (e.g. 5000 AU) and a lower cavity density will measure more J, H, and Ks-band flux, conversely higher cavity density will emit less J, H, and Ks-band flux due to extinction. In addition to possible variations in cavity density, the cavity density is not likely constant as we assume. It could be decreasing as a function of distance from the central source, or vary as function of angle within the cavity to reflect high and low velocity outflow regions \citep{whitney2003a}. Limb brightened cavities are detected in low velocity CO emission and the central jet is detected as high velocity emission \citep{bachiller1995}. We do not explore these possible cavity density variations.

Despite some of the obvious limitations of the images only qualitatively describing our observations, the physical parameters derived from modeling are realistic and conform to previous work \citep[e.g][]{bachiller1995, olinger1999, looney2003, wkwon2006}. In areas where our results do not conform, there are reasons and new insights for the discrepancies highlighted by our data. A caveat that we must mention is that the best fit envelope radius we quote may not be a constraint in reality due to the SED only being taken at a 1000 AU aperture. We regard the distance the scattered light extends as a better measurement of envelope radius, though this value is probably a lower limit because cavity density may vary inversely with radius. The most realistic constraint of envelope radius would be a high resolution extinction map derived from IRAC data; however, such an analysis this is beyond the scope of this paper. For all the sources, except IRS3A, we were able to match a model to the data using our technique of SED fitting and image matching (Table 4). In the following subsections, we discuss each individually. 

\subsection {L1448 IRS 2}

In many ways, L1448 IRS 2 is the most ideal source in the field. It has a curved outflow cavity in which both the red and blue-shifted sides are clearly detected (see Figures 3 and 9a). Also, at first glance, it appears to have evolved in relative isolation compared to the interacting outflows of IRS3A, IRS 3B, and L1448-mm \citep{wkwon2006, olinger2006}. Although \citet{volg2004} detect IRS 2 as a compact binary source, our IRAC data and the MIPS data from the c2d project, do not detect any obvious indicator of its binary nature. The companion lies on the PA ($\sim138^o$) \citep{wolf2000} of the outflow from IRS 2 separated by $\sim$2500 AU ($\sim10^{\prime\prime}$). We will refer to the primary source as IRS 2A and the companion as IRS 2B. We are still able to qualitatively fit an image to IRS 2A as IRS 2B does not appear to affect the morphology of the IRS 2A in our data.

As shown in Figures 9a and 10, the modeled image approximately defines the same cavity boundaries as our data but quantitative differences are evident. First, there is more scattered light emission in the model than the data at 3.6$\mu$m and only one cavity boundary is well defined in the data. The other boundary is indeterminate at 3.6 $\mu$m. At wavelengths longer than 3.6 $\mu$m, the flux in the middle of the cavity dominates the features. However, at 5.8 and 8.0$\mu$m the red shifted side of the cavity shows limb brightening. Also, we observe the cavity flux increasing with wavelength, while the models predict that the cavity flux will fall at longer wavelengths, due to properties of the dust grain model. As discussed previously, incorrectly assumed dust properties may account for the wavelength dependent discrepancies. Also, H$_2$ and PAH emission are not likely to be present in the cavity as IRS spectra do not show evidence of emission lines (IRS Disk Team, private communication). It is also important to note that the emission peaks in each band fall approximately on the PA of the outflow from IRS 2, and the annulus used to construct Figure 10 is coincident with the location of IRS 2B.

Based on the compared images and SED fit for IRS 2A, the inclination and opening angles are 57$^o\pm_8^{6}$ and 20$^o\pm5$ respectively. Our data show that scattered light extends $\sim7000$ AU on the blue-shifted side and $\sim5750$ AU on the red-shifted side. The model image shows scattered light extending to 5000 AU on the red and blue-shifted sides of the cavity. It is unclear whether or not the scattered light will end at the edge of the envelope as it does in the models.

As stated above, IRS 2 is a compact binary; our data only resolve the cavity of IRS 2A. Our IRAC observations could have detected another cavity associated with IRS 2B, which would show the direction of its outflow. The outflow direction of IRS 2B is important as uncertainty remains as to the source of the eastern outflow, HH195E. \citet{bally1997} infers that the source of HH195E is L1448 IRS 1, a Class I source, because of an apparent east-west outflow centered at its location. However, \citet{wolf2000} suggest that another undetected source might be responsible for this outflow. The lack of an \textit{observed} cavity associated with IRS 2B, does not exclude an outflow source from being present. There is a Class 0 source in the region for which we do not clearly detect a cavity, L1448 NW also named L1448 IRS 3C \citep{terebey1997, looney2000}. It seems probable that only high resolution mapping of molecular outflow tracers will fully disentangle these outflows.

\subsection{L1448 IRS 3B}
As Figures 1 and 4 show, IRS 3B is a confused source, lying in close proximity ($\sim7^{\prime\prime}$) to IRS 3A, its binary companion. In addition to being a confused source, the cavity is not symmetric about the outflow PA ($\sim105^o$) \citep{wkwon2006}. In an ideal situation, the cavity would be symmetric about the PA of molecular outflows. With this PA as the axis of symmetry, the southern angle of the cavity is greater than the northern angle, see Figure 4. Only one side of the cavity is clearly visible in our data, making the asymmetry and PA difficult to confirm. However, our data support a $\sim105^o$ PA because HH objects are detected southeast and northwest of IRS 3B on this PA.  

Due to this asymmetry, no model fills the complete extent covered by the observed cavity; models match the northern edge of the cavity, but not the southern edge. One possible reason for this asymmetry is that IRS 3B is being affected by the outflow of L1448-mm. As seen in Figure 1, the outflow jet from L1448-mm is curving toward IRS 3B, possibly confusing the outflow or perhaps eroding that side of the envelope causing the cavity to widen. \citet{olinger2006} posit that the L1448-mm outflow eroded the envelope of IRS 3A. It may be possible that this is happening to IRS 3B as well. Another possibility is that the outflow of IRS 3B is precessing \citep{gueth1996, shepard2000, arce2001, lebron2006, ybarra2006}. HH objects are observed on the PA of IRS 3B and others are present off this PA. If these HH objects located off the PA are associated with IRS 3B, a precessing outflow could account for the HH objects on two PAs as well as an asymmetric cavity. A precessing outflow is not something our data can confirm; it merely suggests the possibility. We must note that L1448 NW (IRS 3C), a Class 0 source, is present near IRS 3B. Despite remaining undetected in our data, it appears to be driving an outflow on a PA of $\sim128^o$. This PA is estimated from tracing HH objects which appear to be collimated toward L1448 NW; molecular line observations have not detected an outflow from L1448 NW. Incidentally, there are HH objects present in the c2d data roughly $\sim0.7$ pc distant from L1448 NW which appear to be roughly coincident with the PA of $\sim128^o$. However, the large distance makes a definite source association uncertain. Nevertheless, it is clear that possibility of multiple outflow sources confuses our analysis of the outflows in association with HH objects.

As shown in Figures 9b and 11, the model predicts flux from the red-shifted side of the cavity at levels too low to distinguish. HH objects with strong emission are present in the vicinity of where the red-shifted side would appear. The blue-shifted side is relatively well matched by model predictions in IRAC channels 1, 2, and 3. Despite being asymmetric, the flux emitted by cavity of IRS 3B corresponds better to its best fitting model than the other sources. IRS 3B is also the only source that does not peak in flux at the center of the cavity in any band. Figure 11 shows some evidence of limb brightening in all bands. Even though IRAC matches very well, discrepancies at H and Ks bands are still present. These discrepancies may be explained by incorrect dust properties and cavity density as described in \S4. H$_2$ and PAH emission are not likely to be present in the cavity as IRS spectra do not show evidence of emission lines (IRS Disk Team, private communication).

Even though IRS 3B is not an ideal source to model, as we do not observe both sides of the cavity; the parameters of the model which best describes the data are consistent with previous studies. The inclination and opening angles, 63$^o\pm_6^7$ and 20$^o\pm5$ respectively, agree with parameters derived in \citet{wkwon2006}. In the IRAC data, scattered light extends to $\sim7500$ AU, possibly implying an envelope radius of the same extent. \citet{looney2003} derived an envelope radius of 8000 AU, however, the adopted distance was 320 pc. Correcting this value for our distance (250 pc), the observed envelope radius is 6250 AU. It is important to note that this envelope is shared with IRS 3A, however, each source appears discretely in our data and can be analyzed independently.

\subsection {L1448-mm}
L1448-mm also has a complex morphology that makes modeling somewhat more difficult. In Figure 5, you see that the source has two bright spots of emission. One is directly north, associated with its cavity and the other is directly south. The central source is located adjacent to the northern brightness peak. The separation between the northern and southern peak is $\sim7^{\prime\prime}$ ($\sim1800$ AU) in IRAC and MIPS channel 1, see Figure 13. Only the northern peak is observed in Ks-band. These data suggest that L1448-mm may be binary in nature. \citet{bachiller1995} mapped L1448-mm at $\lambda =$ 2.7mm continuum with a resolution of 3$^{\prime\prime}$; the data show that L1448-mm is extended along the molecular outflow PA ($\sim159^o$), and \citet{volg2004} detect three distinct sources at $\lambda =$ 2.7mm. The conclusion of \citet{volg2004} was that L1448-mm is a multiple system. Our data support this conclusion, detecting one companion source in our IRAC and MIPS data. We will refer to the northern source, as L1448-mm A and the southern source as L1448-mm B. Only L1448-mm A is associated with the scattered light cavity.

Despite this complex morphology, the scattered light image from L1448-mm A is approximately matched by the model, see Figures 9c and 12. The model matches the data best in IRAC channel 1. In IRAC channels 2, 3, and 4, the flux of the model is too low as compared to the data. However, aside from the indentation on the west side of L1448-mm A, the models approximately match the cavity boundaries of the data in each channel. The data in Figure 12 may show evidence of more scattering at the cavity center, as the flux approximately matches the boundaries predicted by the models but peaks in the middle of the cavity. Notice that the peak flux also lies on the PA of the outflow. Also, some of the wavelength dependent inconsistencies could be the result of improperly assumed dust properties and/or cavity density as discussed previously. No obvious emission lines are present in this source in the wavelengths covered by IRS (IRS Disk Team, private communication).

Although the models do describe the general morphology, the models were not able to completely match the details of L1448-mm A. The northern side of the cavity is approximately matched by the model; however, there is no southern cavity in our data, only L1448-mm B. The brightness of source B could be masking the southern side of the cavity. Also, the scattered light in the model does not extend as far as in the data. The data appear to show scattered light out to $\sim9500$ to 8500 AU from source A. There is a jet present in this region and the emission extending out to 9500 AU could be light scattering on the ambient medium. Also, the limb brightened cavity walls observed in CO J $=1\rightarrow0$ by \citet{bachiller1995} extend $\sim8800$ AU on the northern side and $\sim10000$ AU on the southern side.

The inclination and opening angles of the best matched model, 49$^o\pm8$ and 15$^o\pm5$ respectively, are different, but not extraordinarily so, from the inclination angle of $\sim70^o$ and opening angle of 22.5$^o$ derived from molecular outflow observations \citep{bachiller1995}. \citet{girart2001} derives nearly the same result with a different method. A cavity image using the opening angle and inclinations from \citet{bachiller1995} or \citet{girart2001} could approximately match the data; however, a model SED with these parameters was not fit at the 90$\%$ confidence level. 

The fact that L1448-mm is a multiple system composed of at least 3 sources, possibly 4 \citep{volg2004}, complicates the derivation of physical parameters. Because the companion source is so close in proximity to L1448-mm, our SED measurements are slightly contaminated with flux from L1448-mm B, adding uncertainty to our results. In \S3 it is mentioned that flux increases with as the inclination gets smaller. The additional flux from source B source is likely the reason for fitting smaller inclination angles than \citet{bachiller1995} and \citet{girart2001}.

\subsection{L1448 IRS 3A}

Despite our success with the other sources, L1448 IRS 3A was not constrained at the 90$\%$ confidence level by our SED fitting. This however, is not unexpected because IRS 3A is a Class I source with an envelope mass of $\sim0.09 - 0.29 M_{\sun}$\citep{looney2000, olinger2006}. Our model grid does not attempt to model the Class I stage envelope masses or opening angles. Additionally, the disk properties of a Class I source should have different model input parameters than a Class 0 source. A Class I source is dominated by the disk more than the remnant envelope \citep{looney2000}.

Despite not being constrained by modeling, some general statements can be made from the data. The peak emission in IRAC channel 1 is very near the central source location. Because the source is thought to lie at an inclination of approximately 90$^o$ \citep{wkwon2006}, an emission peak within a few hundred AU from the central source could be evidence of an envelope that is nearly dissipated and a wide opening angle. \citet{olinger2006} posit that the envelope may have been eroded by the outflow from L1448-mm. This hypothesis is supported by the apparent lack of scattered light on the southern side of the source, see Figure 4. If the envelope has been eroded away on the side facing L1448-mm, there would be no scattered light on that side of the source because the cavity in the envelope is the primary scattering surface \citep{whitney2003a}. 

Additionally, we can look for HH objects on the PA of the molecular outflows. In the case of IRS3A, we find emission from HH objects to the north and south of the source that lie on the PA ($\sim155^o$) \citep{wkwon2006}, see Figure 1. The HH objects to the south appear to be collimated toward IRS 3A with conical symmetry. This is probably not scattered light emission because the appearance of scattered light in our data is smooth and continuous while these objects appear ``lumpy.'' The HH objects that appear associated with IRS 3A are all within 50$^{\prime\prime}$. 

\subsection{Opening angle and evolutionary state}
We stated in \S1 that previous work posit that the cavity opening angle will increase as the protostar ages \citep{shu1987, padgett1999, arce2004, arce2006}. Class I protostars modeled by \citet{eisner2005} have opening angles of 25 - 28$^o$ and \citet{terebey2006} constrains the opening angle of TMC-1, a Class I protostar, to be 40$^o\pm$5. Additionally, a small \textit{Hubble Space Telescope} (HST) survey of near edge-on Class I objects in Taurus performed by \citet{padgett1999} measures apparent opening angles $> 30^o$ for all but one source. However, in measuring these opening angles the authors do not account for inclination effects which could account for the outlier. In their sample, they find that opening angle is inversely proportional to circumstellar mass. \citet{stark2006} model these sources in Taurus and obtain fits for opening angles of 30$^o$ (DG Tau B), 25$^o$ (IRAS 4248+2612, 04301+2247), and 20$^o$ (IRAS 04016+2610, CoKu Tau/1).

Our models of the Class 0 sources in L1448 conform to opening angles of 20$^o$ and 15$^o$. If the opening angle does indeed widen with age, there may be a small range of opening angles for which the transition from Class 0 to Class I occurs. With the data to date, we suggest that this transition range is probably between 20$^o$ to 30$^o$. This result, inferred from infrared imaging of cavities, is consistent with studies of outflow cavities at millimeter wavelengths \citep[e.g][]{arce2006}. \citet{arce2006} observes that opening angles of Class 0 sources to be less than 27$\fdg$5, and Class I opening angles are wider. Also, \citet{arce2006} propose an empirical model for the outflow-envelope interaction which causes the opening angle to widen with age. Though our dataset small, it sets the ground work for a more comprehensive study of opening angles observed from cavities in IRAC wavelengths.

\section{Conclusions}

We have demonstrated that the general rule of thumb ``Class 0 sources
are invisible at $\lambda < 10 \mu m$'' is incorrect.  New advances in
instrumentation allow us to probe into the circumstellar envelope by
imaging the scattered light from the central protostar that is escaping
out the cavity, carved by the powerful molecular outflows.
In fact, the scattered light emission is brightest in the 
IRAC bands due to the lower extinction.
These images impose, for the first time, constraints on the central
protostar and environment in a new and exciting way.
In particular, we have shown that the general morphologies of Class 0 objects
(i.e. circumstellar envelope with a cavity, circumstellar disk, and
embedded protostar)
are consistent with our scattered light images.
The ability to which the
models match the data depends largely on the circumstellar environment
of the protostar. If the protostar has a very ideal morphology
(i.e. spherical and axial symmetry), it is possible to well model
the source. Complex morphologies e.g. IRS 3B and L1448-mm are more
problematic. Even so, an ideal model that approximates the structure
of an non-ideal source provides important constraints on its
circumstellar structure, mainly opening angle and inclination angle,
as well as providing loose constraints on the source parameters.

We are able to fit an opening angle and inclination angle to our sources
using both images and SED fits. For L1448 IRS 2, the source that best
exhibits a bipolar cometary nebula, our best fit inclination angle is
57$^o\pm_8^{6}$ and the opening angle is 20$^o\pm5$. L1448 IRS 3B,
a source which may have an asymmetric cavity, has a best fit inclination
angle of 63$^o\pm_6^7$ and the opening angle is 20$^o\pm5$. L1448-mm,
the source that drives one of the most prominent examples of molecular
outflows, has a best fit inclination angle of 49$^o\pm8$ and the
opening angle is 15$^o\pm5$.

Deriving the outflow opening and inclination angles are crucial
to understand the sources.
There is a probable link between the age of the source and opening angle;
observations and models suggest that the opening angle widens with
age. The inclination angle can completely change the appearance
of the cavity, interacting with the apparent opening angle. 
In other words, one must derive the
true opening angle of an embedded source with the
inclination to decouple the interdependency of the geometry and extinction.
We assert that the opening
angle of a protostar can be indicative of its evolutionary state based on
our modeling of Class 0 sources in L1448 and previous studies of outflow
cavities. Also, comparing the inclination angle derived from modeling
to an inclination angle derived from molecular outflow observations is
an important baseline. A possible constraint on envelope radius could
be the distance the scattered light extends from the source; however,
this distance is likely a lower limit.

\textit{Spitzer} gives us a very unique and detailed glimpse of circumstellar structure around protostars. At the same time, however, it reveals that the circumstellar structure and formation environments are generally far from being idealized cases. Analysis of models versus data does provide important constraints of the circumstellar structure of protostars, but we may still be missing important details. In the near future, we may be able to more quantitatively constrain the circumstellar structure. New three-dimensional radiative transfer codes \citep[e.g.][]{whitney2005, indeb2006} may enable us to model even the most unique circumstellar environments. However, with new radiative transfer codes the multitude of parameters increases as does the time required to model a comprehensive grid. Observational programs must be able to constrain input parameters to reduce complexity. Future instruments, e.g. \textit{James Webb Space Telescope}, Atacama Large Millimeter Array, will be able to provide astronomers with highly sensitive and detailed observations of protostars, building upon the foundation that \textit{Spitzer} has laid. 

\acknowledgements
The authors wish to thank B. A. Whitney for providing her model and
helpful discussions regarding its use, Robert Gruendl for
helpful discussions and assistance with the figures, and 
Lee Hartmann and Nuria Calvet for stimulating discussions
during the final stages of this project. We also would like to thank
the anonymous referee for comments which improved the relevance of this paper.
This publication makes use of data from the COordinated Molecular Probe Line 
Extinction Thermal Emission (COMPLETE) Survey of Star Forming Regions, the \textit{cores2disks}
legacy program for the \textit{Spitzer Space Telescope}. This publication
makes use of data products from the Two Micron All Sky Survey, which
is a joint project of the University of Massachusetts and the Infrared
Processing and Analysis Center/California Institute of Technology, funded
by the National Aeronautics and Space Administration and the National
Science Foundation. J.T., L.W.L., M.H., and W.K. acknowledge support from
\textit{Spitzer Space Telescope} grant FCPO. 1264492. L.G.M acknowledges
support from \textit{Spitzer Space Telescope} grant FCPO. 1264478.

\bibliographystyle{apj}
\bibliography{ms}

\clearpage

\begin{deluxetable}{lllc}
\tablewidth{0pt}
\tablecaption{Source List and Extinction\label{source}}
\tablehead{
  \colhead{Source} & \colhead{Coordinates (J2000.0)} & \colhead{} & \colhead{A$_{v}$\tablenotemark{a}}\\
  \colhead{} & \colhead{Right Ascension} & \colhead{Declination} & \colhead{}\\
  }
\startdata
L1448 IRS 3A & 03h 25m 36.532s & +30$^o$45m 21.35s & 4.57\\
L1448 IRS 3B & 03h 25m 36.339s & +30$^o$45m 14.94s & 4.57\\
L1448-mm    & 03h 25m 38.8s   & +30$^o$44m 05s    & 5.4 \\
L1448 IRS 2  & 03h 25m 22.4s   & +30$^o$45m 12s    & 4.37
\enddata

\tablenotetext{a}{Derived from \citet{complete2006}.}
\end{deluxetable}

\begin{deluxetable}{lcccccccc}
\tablewidth{0pt}
\tabletypesize{\small}
\tablecaption{Photometry\label{photo}}
\tablehead{
  \colhead{Source} & \colhead{Aperture} & \colhead{F$_{J}$} & \colhead{F$_{H}$} & \colhead{F$_{Ks}$} & 
  \colhead{F$_{3.6}$} & \colhead{F$_{4.5}$} & \colhead{F$_{5.8}$} & \colhead{F$_{8.0}$}\\
  \colhead{} & \colhead{(AU)} & \colhead{(mJy)} & \colhead{(mJy)} & \colhead{(mJy)} & 
  \colhead{(mJy)} & \colhead{(mJy)} & \colhead{(mJy)} & \colhead{(mJy)}\\
  }
\startdata
L1448 IRS 3A & 1000 & 0.172 & 0.293 & 0.750 & 2.54 & 16.6 & 51.6 & 96.6\\
L1448 IRS 3B & 1000 & ... & ... & 0.380 & 2.44 & 6.85 & 9.80 & 11.7\\
L1448-mm A   & 1000 & ... & 0.144 & 1.07 & 4.80 & 15.5 & 29.7 & 65.1\\
L1448 IRS 2A  & 1000 & ... & ... & 0.117 & 1.15 & 5.38 & 10.2 & 15.8\\
\enddata
\tablecomments{Photometric measurements have been corrected for extinction. J, H, and Ks, flux values have uncertainties of 15$\%$, 11$\%$, and 12$\%$ respectively. IRAC flux values all have uncertainties of 15$\%.$}
\end{deluxetable}

\begin{deluxetable}{llc}
\tablewidth{0pt}
\tablecaption{Model grid parameters\label{param}}
\tablehead{
  \colhead{Parameter} & \colhead{Description} & \colhead{Value}\\
  }
\startdata
R$_{*}$(\rsun) & Stellar radius & 2.09, 3, 3.62, 4.18 4.67, 5.12, 5.53, 5.91\\
T$_{*}$(K) & Stellar temperature & 4000\\
L$_{*}$(L$_{\sun}$) & System luminosity\tablenotemark{a} & 2.11, 2.84, 3.65, 4.57, 5.51, 6.48, 7.45, 8.42\\
M$_{*}$(M$_{\sun}$) & Stellar mass & 0.5\\
M$_{disk}$(M$_{\sun}$) & Disk mass & 0.05\\
$\alpha$ & Disk radial density exponent & 2.25\\
$\beta$ & Disk scale height exponent & 1.25\\
$\dot{M}_{disk}$(M$_{\sun}$ $yr^{-1}$) &  Disk accretion rate & 2 $\times 10^{-7}$\\
R$_{disk,min}$(R$_{sub}$)\tablenotemark{b} & Disk inner radius & 1\\
R$_{disk,max}$(AU) & Disk outer radius & 50\\
R$_{c}$(AU) & Centrifugal radius & 3\\
R$_{env,min}$(R$_{sub}$) & Envelope inner radius & 3\\
R$_{env,max}$(AU) & Envelope outer radius & 5000, 6500, 8000\\
M$_{env}$\tablenotemark{c} & Envelope mass & 0.25, 0.5, 0.75, 1.0, 1.25, 1.5, 1.75, 2.0\\
$\theta_{open}$ & Opening angle & 10$^o$, 15$^o$, 20$^o$, 25$^o$, 30$^o$\\
$\theta_{inc}$ & Inclination angle & cos \textit{i} = 0.05, 0.15, ..., 0.95\\
$\rho_{c}$(g cm$^{-3}$) & Cavity density & 3.34 $\times$ 10$^{-19}$
\enddata
\tablenotetext{a}{The system luminosity is not an explicitly varied parameter, but it is greatly dependent on R$_{*}$. }
\tablenotetext{b}{R$_{sub}$ is the dust destruction (sublimation) radius. This is calculated internally by the model and will vary with the stellar radius/luminosity.}
\tablenotetext{c}{The envelope mass is set by the mass infall rate. The mass infall rate for a given envelope mass changes with different envelope radii. For brevity, we only list the approximate envelope masses.}
\end{deluxetable}

\begin{deluxetable}{lccccccc}
\tablewidth{0pt}
\tabletypesize{\small}
\tablecaption{Best Fit Parameters}
\label{fit_01}
\tablehead{
  \colhead{Source} & \colhead{Luminosity}\tablenotemark{a} & \colhead{Stellar}\tablenotemark{b} & \colhead{Envelope}\tablenotemark{c} & \colhead{Envelope}\tablenotemark{d} & \colhead{Opening} & \colhead{Inclination} & \colhead{SED}\\
  \colhead{} & \colhead{} & \colhead{Radius} & \colhead{Radius} & \colhead{Mass} & 
  \colhead{Angle} & \colhead{Angle} & \colhead{$\chi_r^{2}$}\\
  \colhead{} & \colhead{(L$_{\sun}$)} & \colhead{(R$_{\sun}$)} & \colhead{(AU)} & \colhead{(M$_{\sun}$)} & 
  \colhead{($^o$)} & \colhead{($^o$)} & \colhead{}\\
  }
\startdata
L1448 IRS2  & 4.6  & 4.2 & 5000$\pm1500$  & 1.50$\pm0.5$ & 20$\pm5$ & 57$\pm_8^{6}$ & 0.374\\
L1448 IRS 3B & 4.6 & 4.2 & 5000$\pm1500$ & 1.00$\pm0.5$ & 20$\pm5$ & 63$\pm_6^7$ & 0.603\\
L1448-mm    & 7.5  & 5.5 & 6500$\pm1500$  & 1.75$\pm0.5$ & 15$\pm5$ & 49$\pm8$ & 1.62
\enddata
\tablenotetext{a}{The value found to best fit the 1.66 - 8.0$\mu$m data; it is likely a lower limit on the bolometric luminosity.}
\tablenotetext{b}{This value depends on assumptions in our model grid and best-fit luminosity.}
\tablenotetext{c}{Envelope radius is weakly constrained by the fits. The error bars represent fit ranges within the model grid which likely underrepresent the total uncertainty.}
\tablenotetext{d}{The uncertainty in envelope mass is tied to the uncertainty in envelope radius. Also if the density power law is different from r$^{-3/2}$ \citep{looney2003} the envelope mass is more uncertain. A fit within the model grid simply fits the optical depth necessary to produce a similar SED for a given luminosity. As seen in Table 5, a multitude of different envelope masses and radii produce an SED fit.}
\end{deluxetable}

\begin{deluxetable}{lccccccc}
\tablewidth{0pt}
\tabletypesize{\footnotesize}
\tablecaption{Results from SED fitting}
\label{fit_02}
\tablehead{
  \colhead{Source} & \colhead{Luminosity} & \colhead{Stellar} & \colhead{Envelope} & \colhead{Envelope} & 
  \colhead{Opening} & \colhead{Inclination} & \colhead{SED }\\
  \colhead{} & \colhead{} & \colhead{Radius} & \colhead{Radius} & \colhead{Mass} & 
  \colhead{Angle} & \colhead{Angle} & \colhead{$\chi_r^{2}$}\\
  \colhead{} & \colhead{(L$_{\sun}$)} & \colhead{(R$_{\sun}$)} & \colhead{(AU)} & \colhead{(M$_{\sun}$)} & 
  \colhead{($^o$)} & \colhead{($^o$)} & \colhead{}\\
  }
\startdata
L1448 IRS 3B & 2.11 & 2.09 & 8000 & 1.25 & 15 &   63  &  1.27\\
& 2.84 & 3 & 8000 & 1.75 & 20 &   63  &  1.42\\
& 4.57 & 4.18 & 5000 & .50 & 10 &   81  &  1.55\\
& 4.57 & 4.18 & 5000 & 1 & 20 &   63  &  0.603\\
& 5.51 & 4.67 & 8000 & 1.25 & 10 &   81  &  1.59\\
& 6.48 & 5.12 & 5000 & 1.75 & 20 &   57  &  1.19\\
& 6.48 & 5.12 & 6500 & 1 & 10 &   76  &  0.555\\
& 6.48 & 5.12 & 6500 & 1.5 & 10 &   49  &  1.29\\
& 6.48 & 5.12 & 6500 & 1.75 & 20 &   70  &  1.84\\
& 6.48 & 5.12 & 8000 & 1.5 & 10 &   63  &  1.41\\
& 6.48 & 5.12 & 8000 & 1.75 & 15 &   81  &  1.95\\
& 6.48 & 5.12 & 8000 & 2 & 15 &   70  &  1.86\\
& 7.45 & 5.53 & 5000 & 1 & 15 &   81  &  1.53\\
& 7.45 & 5.53 & 6500 & 2 & 15 &   57  &  0.909\\
& 8.42 & 5.91 & 8000 & 2 & 15 &   81  &  1.77\\
L1448-mm& 2.11 & 2.09 & 5000 & .75 & 10 &   32  &  1.70\\
& 2.11 & 2.09 & 6500 & .75 & 15 &   57  &  1.20\\
& 2.11 & 2.09 & 6500 & 1.25 & 30 & 63 & 1.79\\
& 2.11 & 2.09 & 8000 & 1.75 & 30 & 63 & 0.822\\
& 2.84 & 3 & 5000 & 2 & 15 &   32  &  1.85\\
& 2.84 & 3 & 8000 & 1.25 & 10 &   41  &  1.97\\
& 3.65& 3.62 & 6500 & 1 & 10 &   41  &  1.91\\
& 3.65 & 3.62 & 6500 & 1 & 15 &   57  &  1.70\\
& 4.57 & 4.18 & 6500 & 1.5 & 20 &   57  &  1.63\\
& 4.57 & 4.18 & 8000 & 1.25 & 10 &   49  &  1.64\\
& 5.51 & 4.67 & 5000 & .75 & 10 &   41  &  1.72\\
& 5.51 & 4.67 & 5000 & 1 & 15 &   49  &  1.82\\
& 5.51 & 4.67 & 8000 & 1.75 & 10 &   41  &  1.95\\
& 7.45 & 5.53 & 6500 & 1.75 & 15 &   49  &  1.62\\
& 8.42 & 5.91 & 5000 & 1.75 & 25 & 57 & 1.21\\
L1448 IRS2 & 2.11 & 2.09 & 5000 & 1.75 & 20 &   49  &  1.97\\
& 2.11 & 2.09 & 6500 & 2 & 30 & 63 & 1.93\\
& 2.11 & 2.09 & 8000 & 1.75 & 15 &   57  &  1.03\\
& 3.65 &  3.62 & 5000 & 1.25 & 15 &   49  &  0.752\\
& 4.57 & 4.18 & 5000 & 1.5 & 20 &   57  &  0.373\\
& 6.48 & 5.12 & 6500 & 2 & 10 &   41  &  1.79
\enddata

\end{deluxetable}

\clearpage

\begin{figure}
\figurenum{1}
\plotone{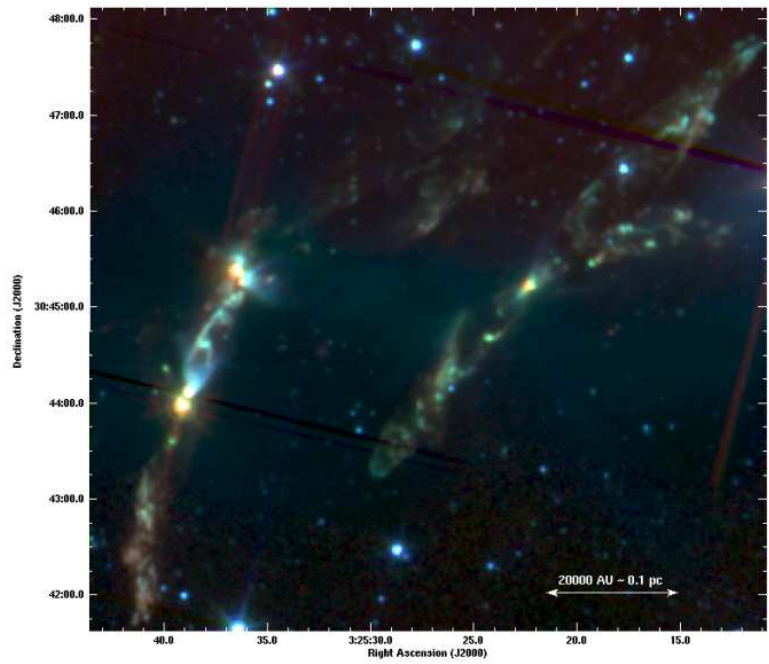}
\caption{IRAC image of L1448, color composite of Channels 1 (blue), 2 (green) and 4 (red). This image is combined with the c2d data for full wavelength coverage across the field of view.}
\end{figure}

\begin{figure}
\figurenum{2}
\plotone{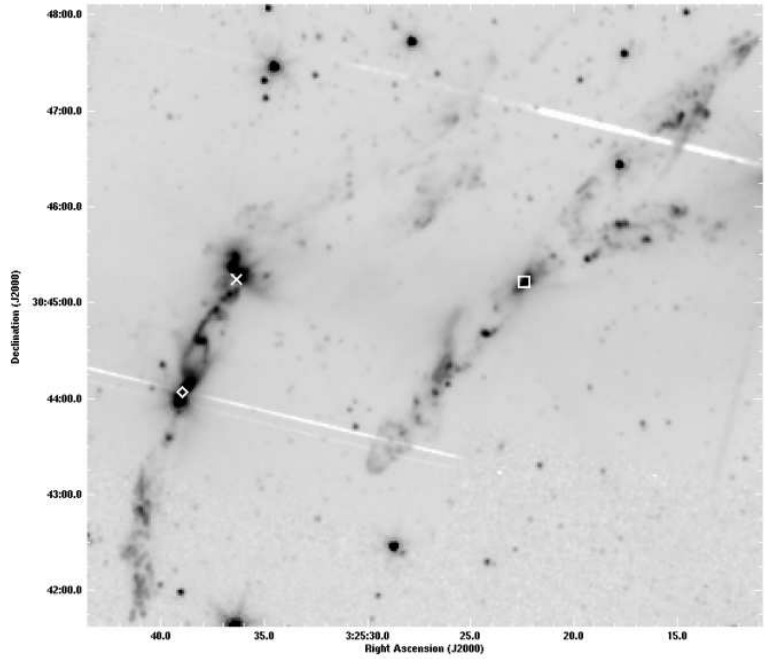}
\caption{IRAC channel 2 image of L1448. Major sources are marked: IRS 2 is marked with an open box, IRS 3 is marked with a X, and L1448-mm is marked with an open diamond. The companion sources are identified in Figures 3, 4 and 5.}
\end{figure}

\begin{figure}
\figurenum{3}
\includegraphics[angle=-90, scale=.8]{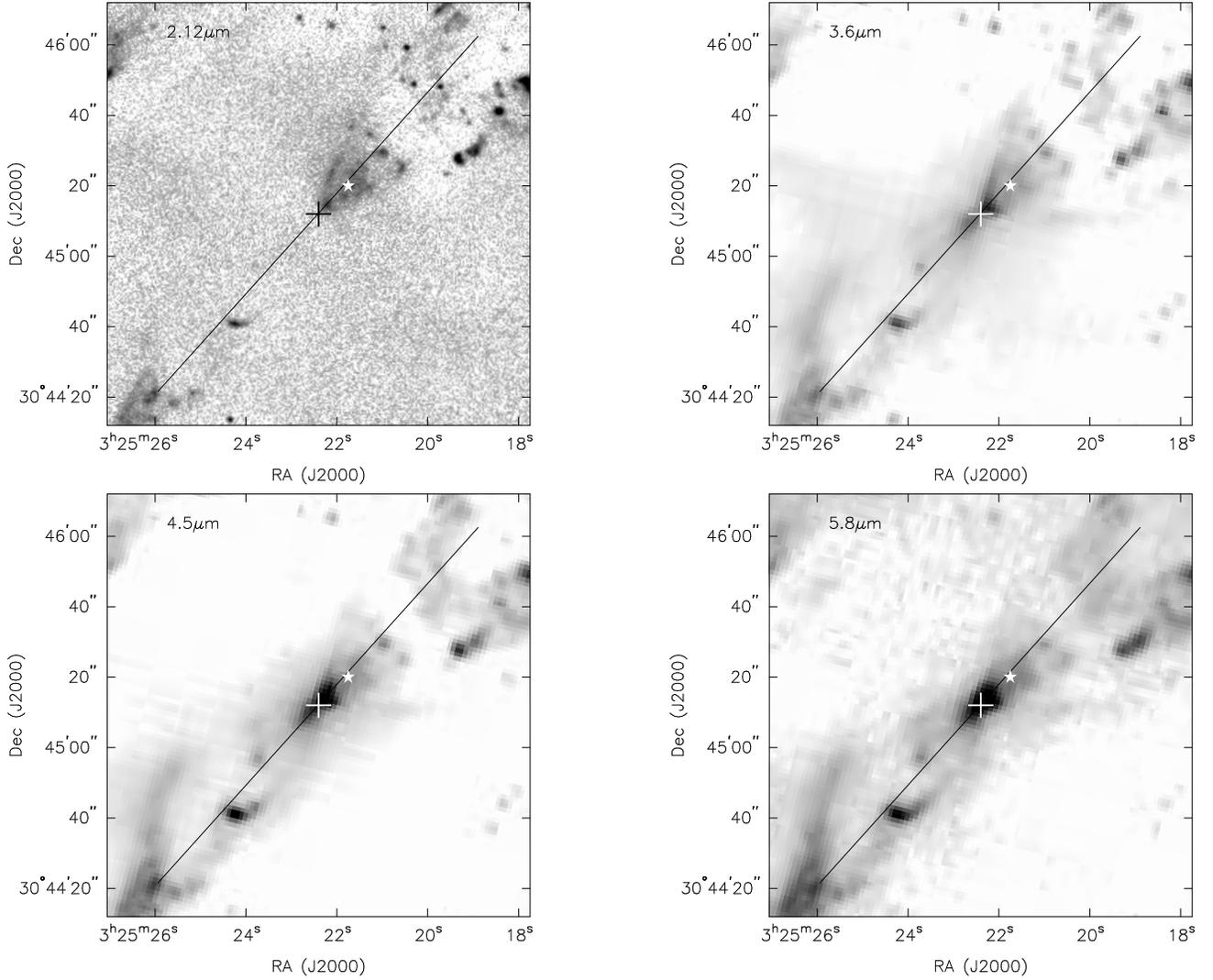}
\caption{Zoomed in images of cavities for IRS 2. Images are 2$^{\prime}$ (30,000 AU) on each side. The solid line is the PA of the CO outflow from IRS 2A from \citet{wolf2000}, the cross is the location of the peak millimeter emission from IRS 2A, and the star is the location of IRS 2B. Note that there is no obvious detection of IRS 2B.}
\end{figure}

\begin{figure}
\figurenum{4}

\includegraphics[angle=-90, scale=.8]{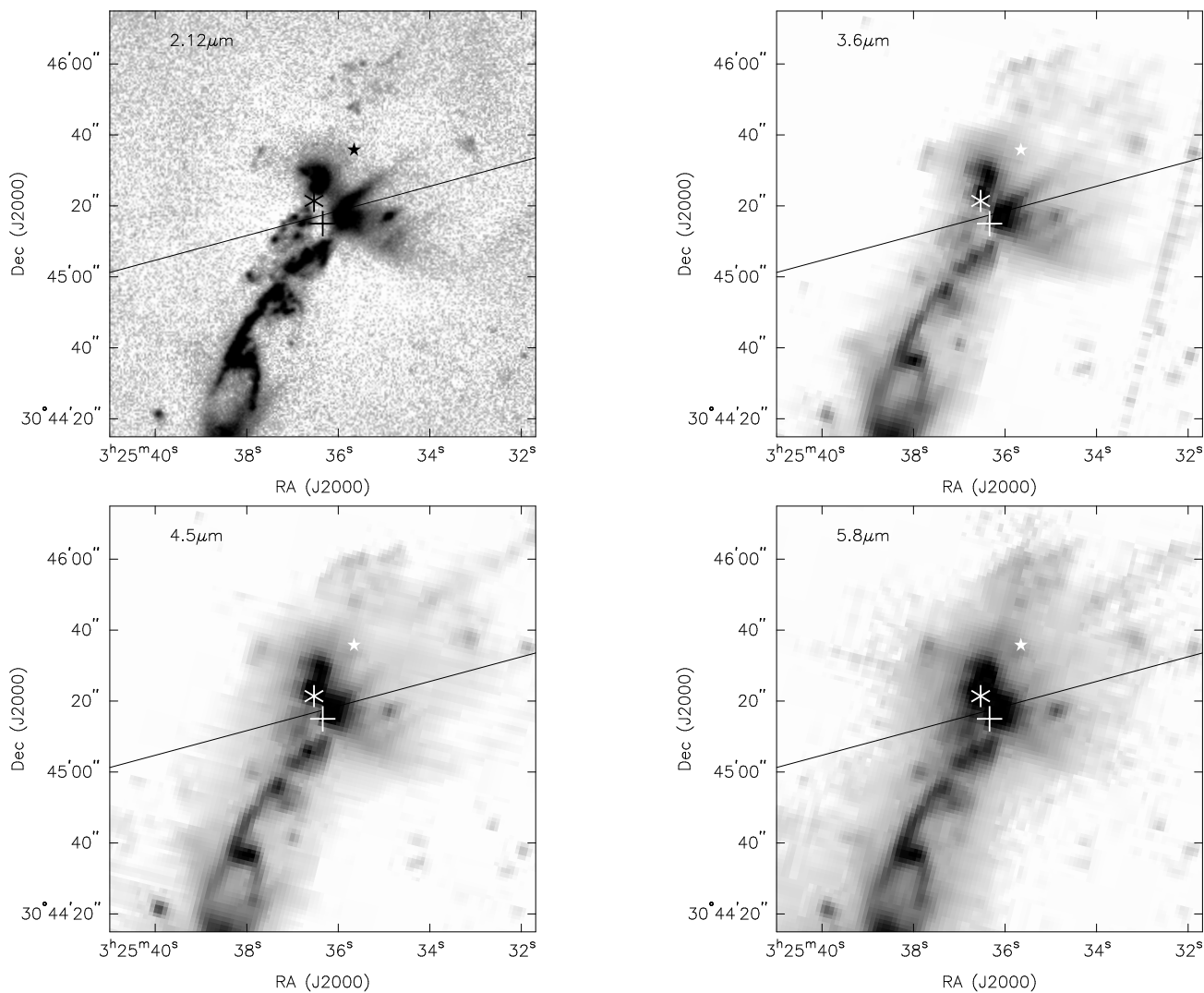}
\caption{Zoomed in images of cavities for IRS 3B. Images are 2$^{\prime}$ 30,000 AU) on each side. The solid line is the PA of the CO outflow from IRS 3B from \citet{wkwon2006}. The cross is the location of the peak millimeter emission from IRS 3B, the asterisk is the location of IRS 3A, and the star is the location of IRS 3C (L1448 NW). Note that the cavity of IRS 3B is not symmetric about the PA of the CO outflow. Also, there is a marginal detection of emission near IRS 3C in IRAC channels 3 and 4.}

\end{figure}

\begin{figure}
\figurenum{5}
\includegraphics[angle=-90, scale=.8]{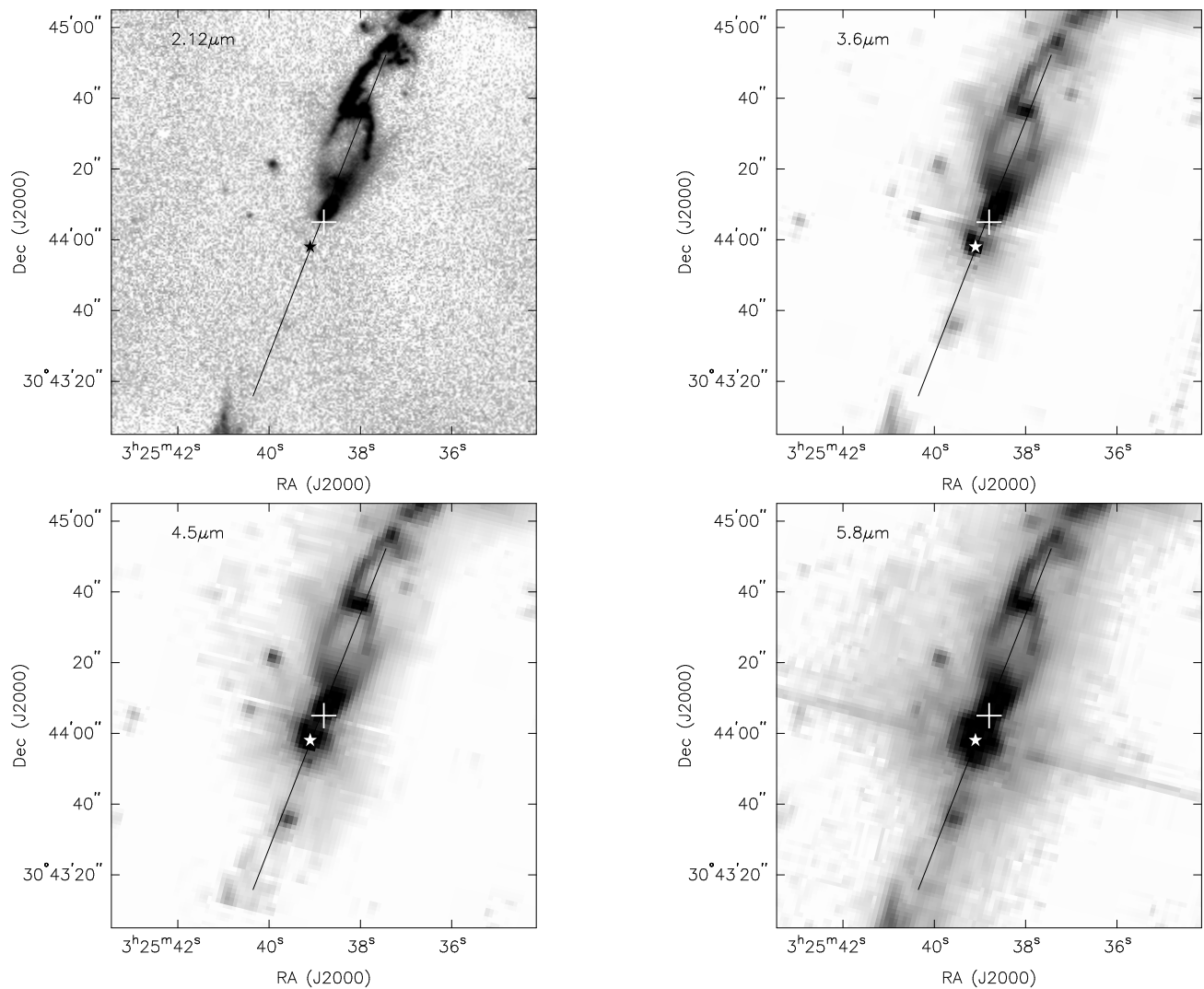}
\caption{Zoomed in images of cavities for L1448-mm. Images are 2$^{\prime}$ (30,000 AU) on each side.The solid line is the PA of the CO outflow from L1448-mm A from \citet{bachiller1995}. The cross is the location of the peak millimeter emission from L1448-mm A, and the star is the location of the peak millimeter emission from L1448-mm B. Note that L1448-mm B is undetected in the Ks-band and becomes more prominent at longer wavelengths.}

\end{figure}

\begin{figure}
\figurenum{6}
\plotone{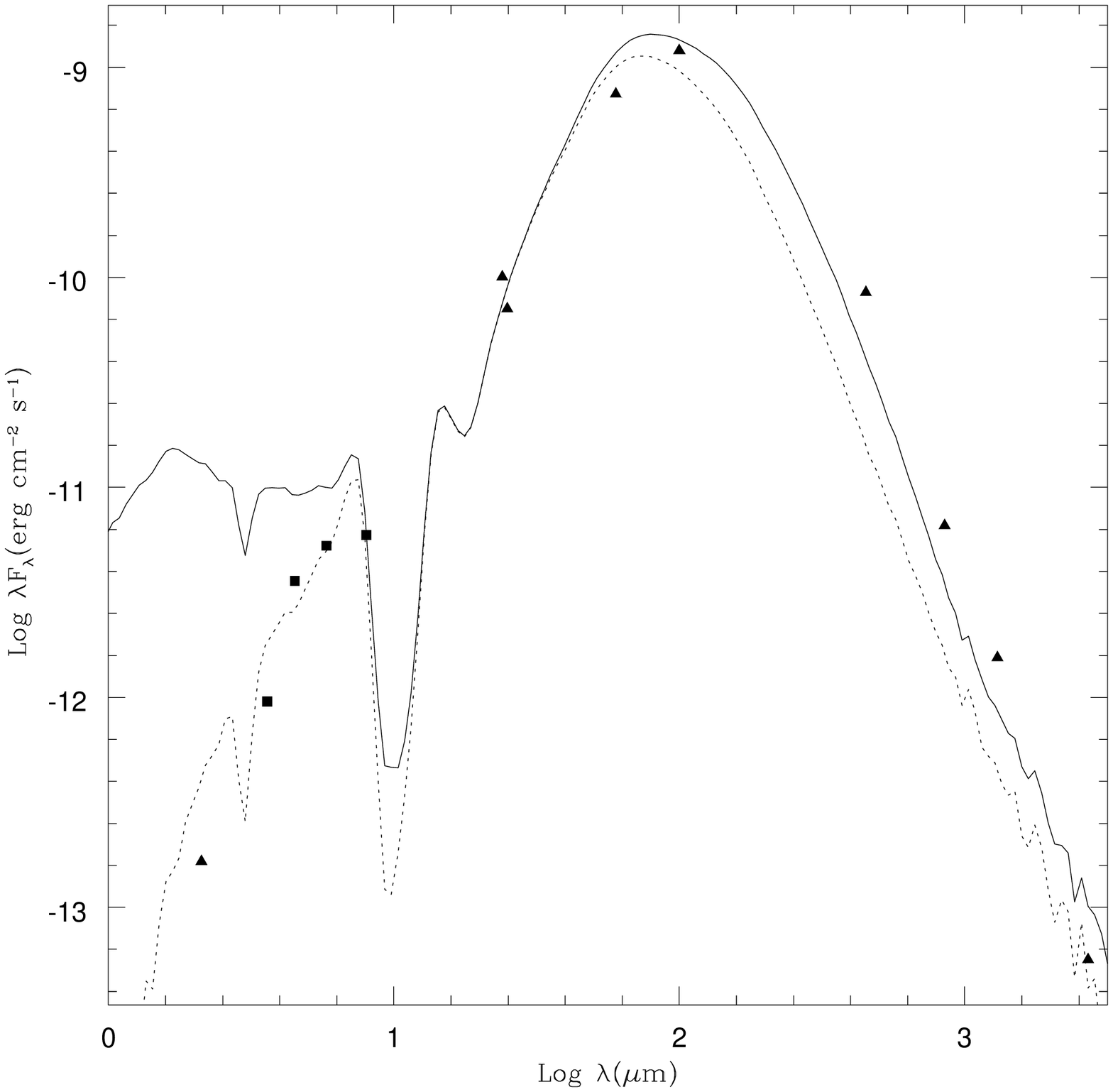}
\caption{SED of of best fit model for IRS 2A at apertures of 1000 (dashed line) and 5000 AU (solid line), plotted with photometry. Data points represented as boxes were included in fitting, triangles were not used in fitting but are included for completeness. Wavelengths greater than 8.0$\mu$m are upper limits, as the aperture sizes are greater than our 1000 AU aperture. IRS 2 is a multiple system and all data except the $\lambda = $ 2.7mm data point include the companion sources.  $\lambda = $ 12, 25, 60, 100, 450, 850, and 1300$\mu$m data points are from \citet{olinger1999}, and the $\lambda = $ 2.7mm data point is from \citet{volg2004}. }
\end{figure}

\begin{figure}
\figurenum{7}
\plotone{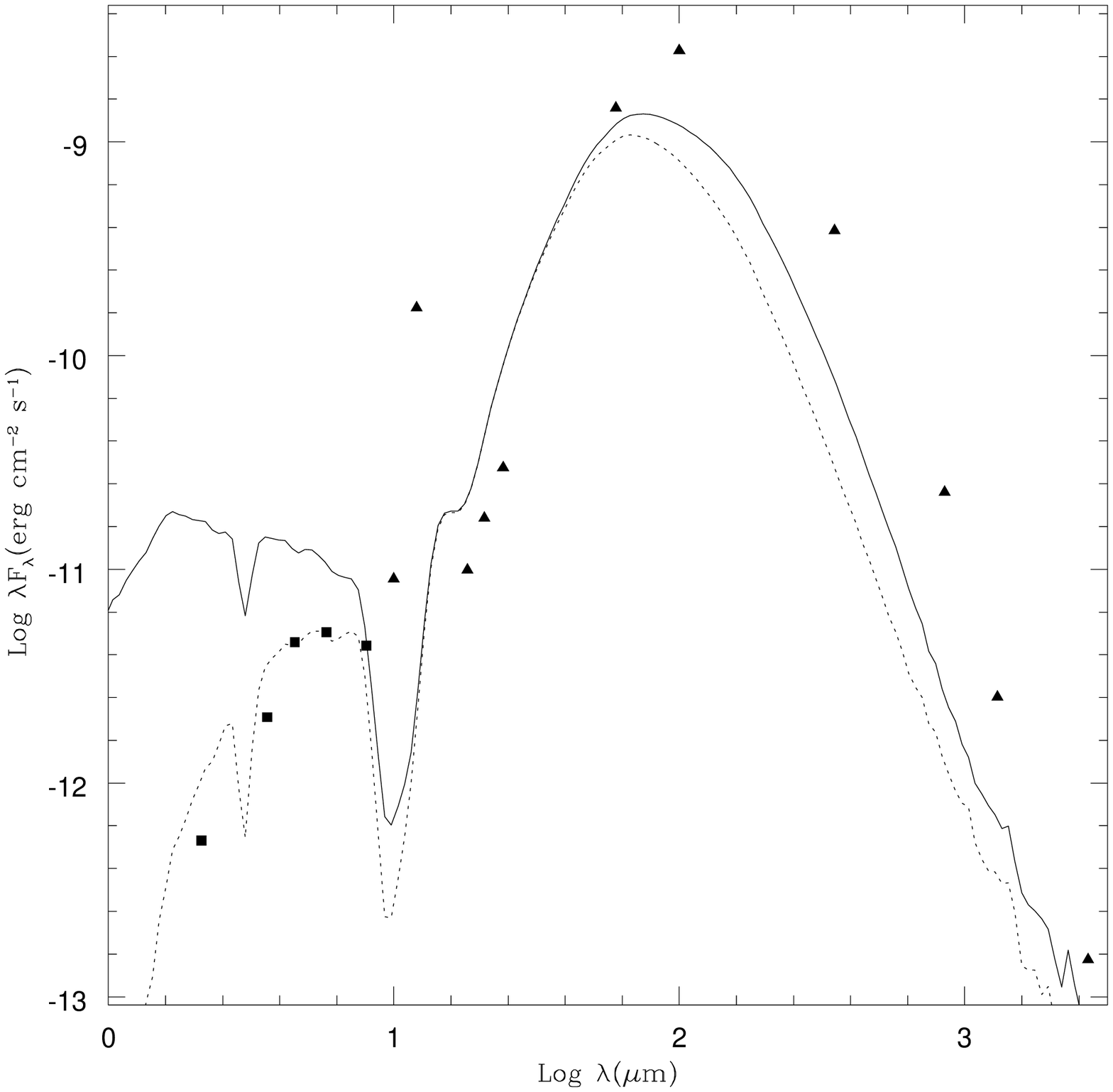}
\caption{SED of of best fit model for IRS 3B at apertures of 1000 (dashed line) and 5000 AU (solid line), plotted with photometry. Data points represented as boxes were included in fitting, triangles were not used in fitting but are included for completeness. Wavelengths greater than 8.0$\mu$m are upper limits, as the aperture sizes are greater than our 1000 AU aperture. Only the $\lambda = $ 850$\mu$m and greater data points resolve IRS 3B from its companion. 10.38, 18.08, 20.75 and 24.13$\mu$m data points are upper limits from \citet{ciardi2003}, \citep{olinger2006} give greater upper limits; $\lambda = $ 12, 25, 60, 100, and 350$\mu$m data points are from \citet{barsony1998}, $\lambda = $ 850 and 1300$\mu$m data points are from \citet{wkwon2006}, and the $\lambda = $2.7mm data point is from \citet{looney2000}.}
\end{figure}

\begin{figure}
\figurenum{8}
\plotone{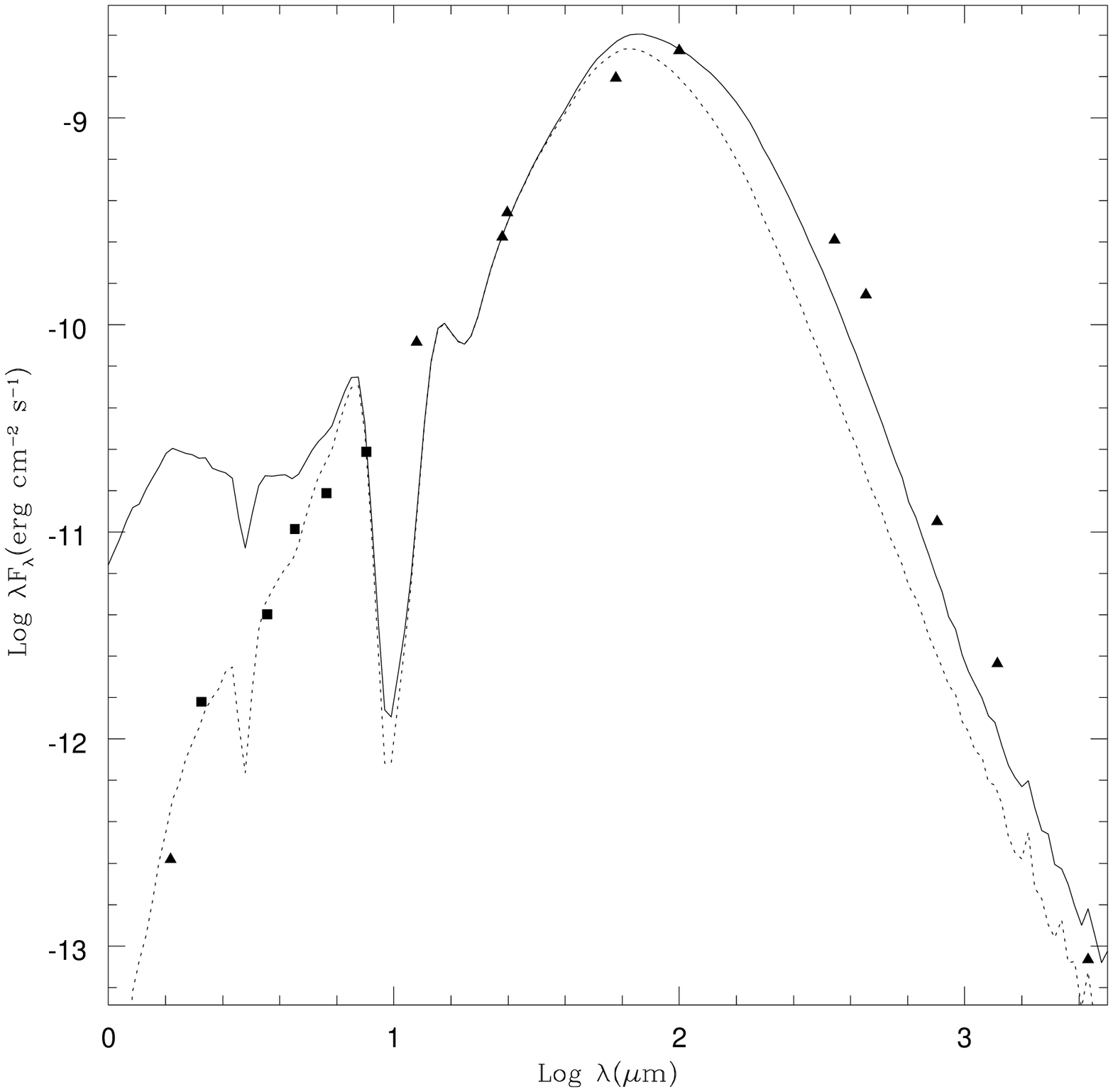}
\caption{SED of of best fit model for L1448-mm A at apertures of 1000 (dashed line) and 5000 AU (solid line), plotted with photometry. Data points represented as boxes were included in fitting, triangles were not used in fitting but are included for completeness. Wavelengths greater than 8.0$\mu$m are upper limits, as the aperture sizes are greater than our 1000 AU aperture. Only the $\lambda = $ 2.7mm data point resolves the source itself, the rest include flux from companion sources. $\lambda = $ 12, 25, 60, 100, 350, 450, 800, and 1300$\mu$m data points are from \citet{barsony1998}, and the $\lambda = $ 2.7mm data point is from \citet{volg2004}. }
\end{figure}

\begin{figure}
\figurenum{9a}
\includegraphics[angle=-90, scale=.8]{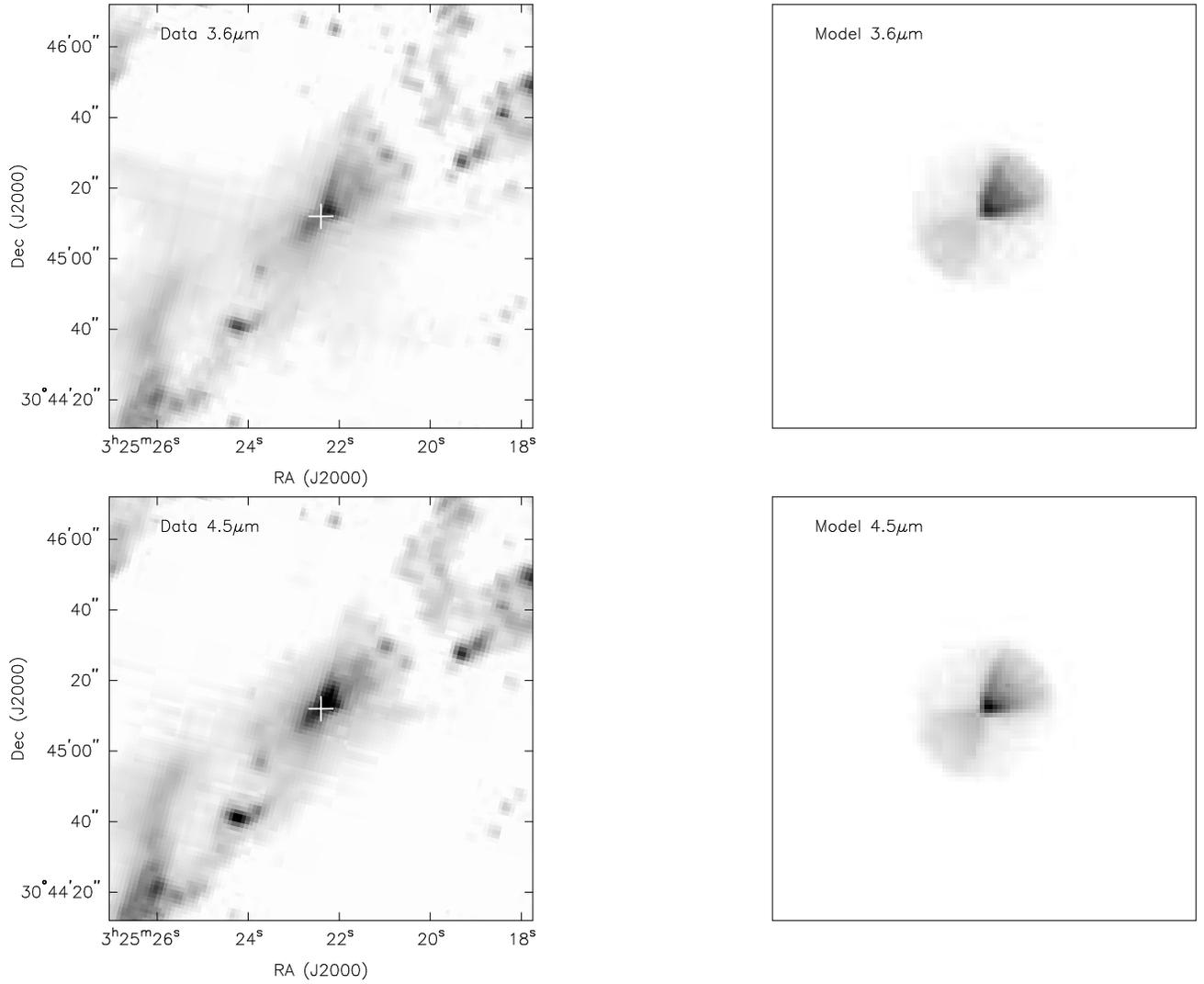}
\caption{Zoomed in images of cavities modeled for each source: (a) IRS 2, (b) IRS 3B, and (c) L1448-mm. Images are 2$^{\prime}$ (30,000 AU) on each side; the primary source being modeled is marked with a cross.}
\end{figure}
\clearpage
\centerline{\includegraphics[angle=-90, scale=.8]{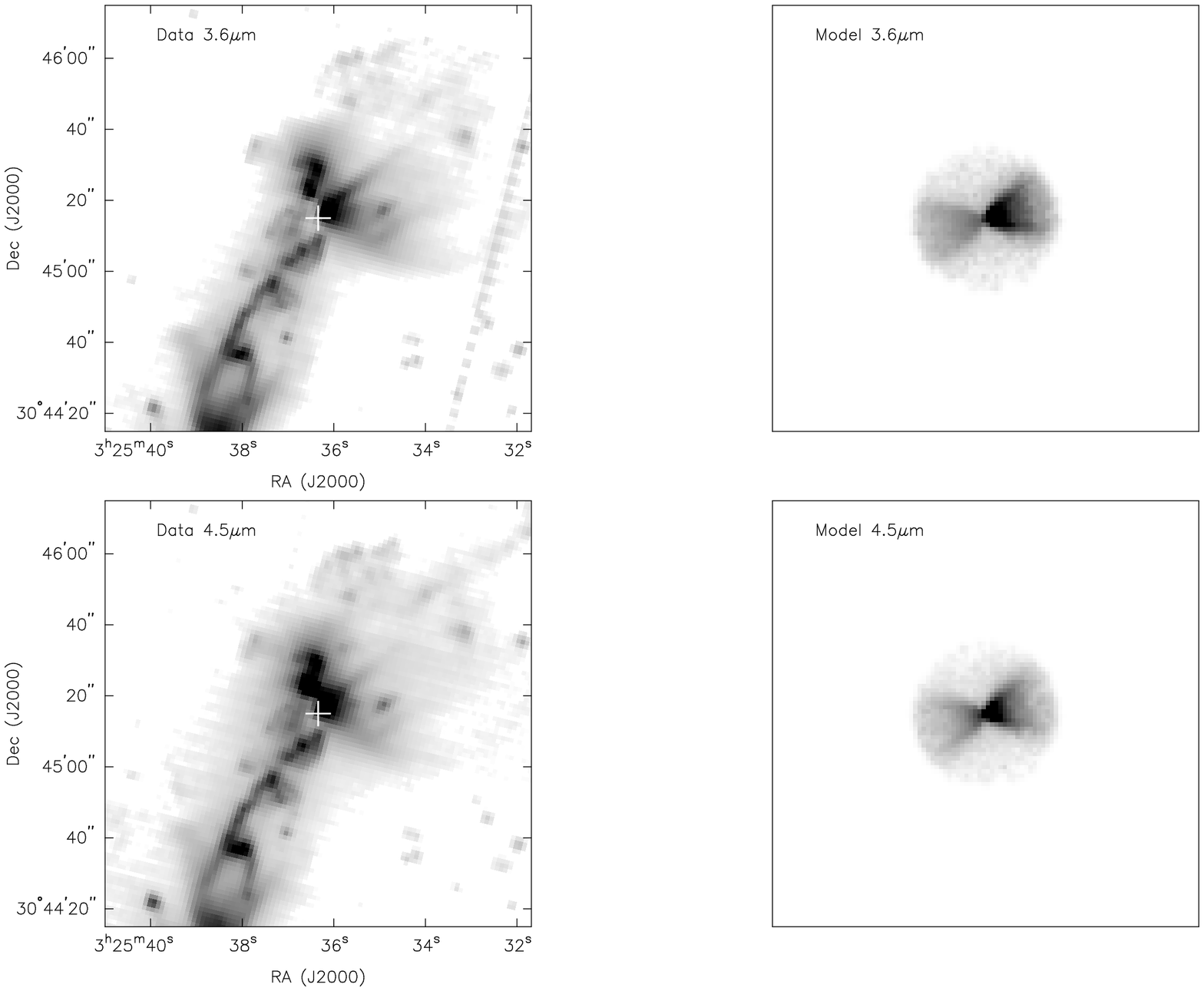}}
\centerline{Fig. 9b. ---}
\clearpage
\centerline{\includegraphics[angle=-90, scale=.8]{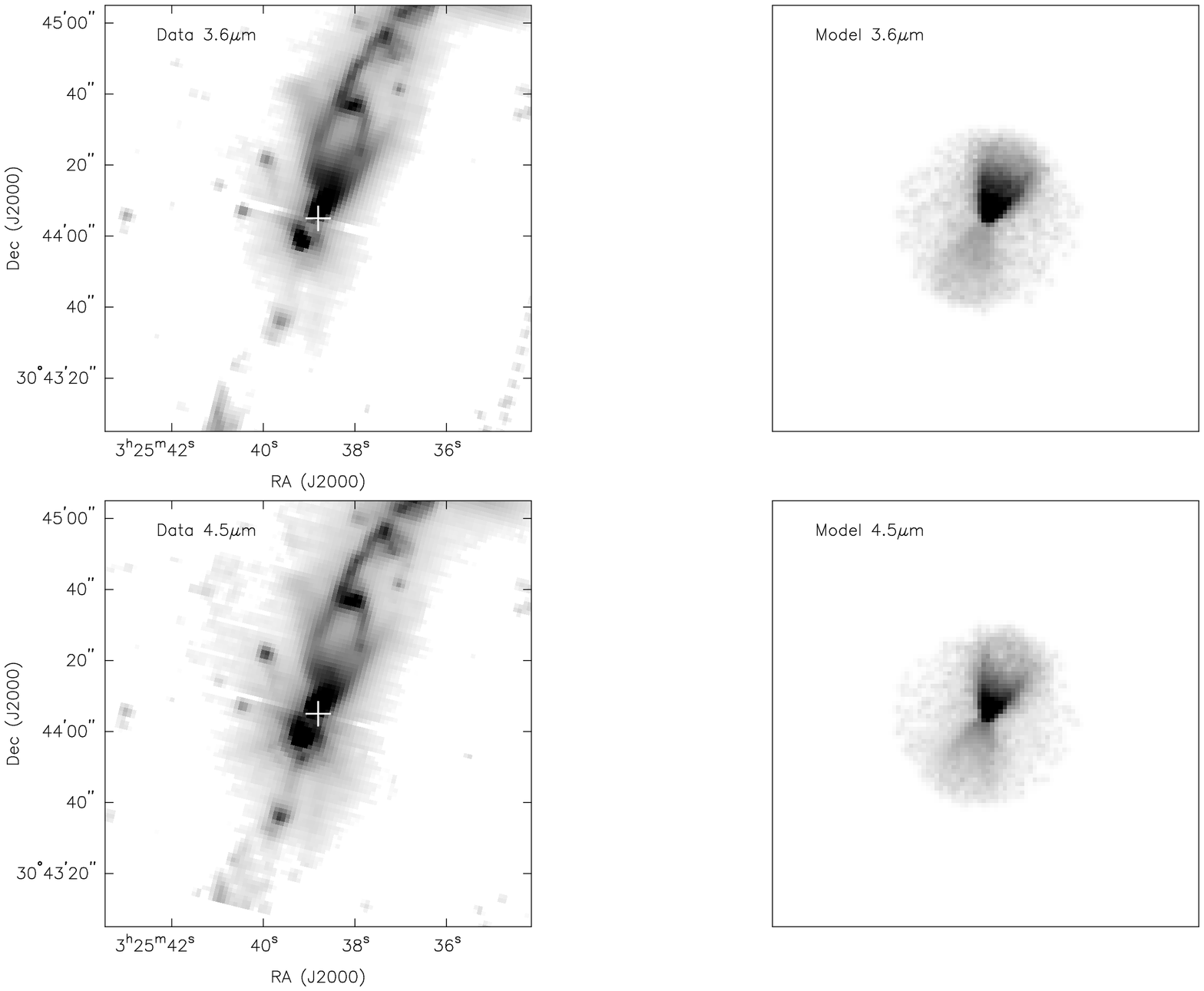}}
\centerline{Fig. 9c. ---}

\begin{figure}
\figurenum{10}
\includegraphics[angle=-90, scale=.65]{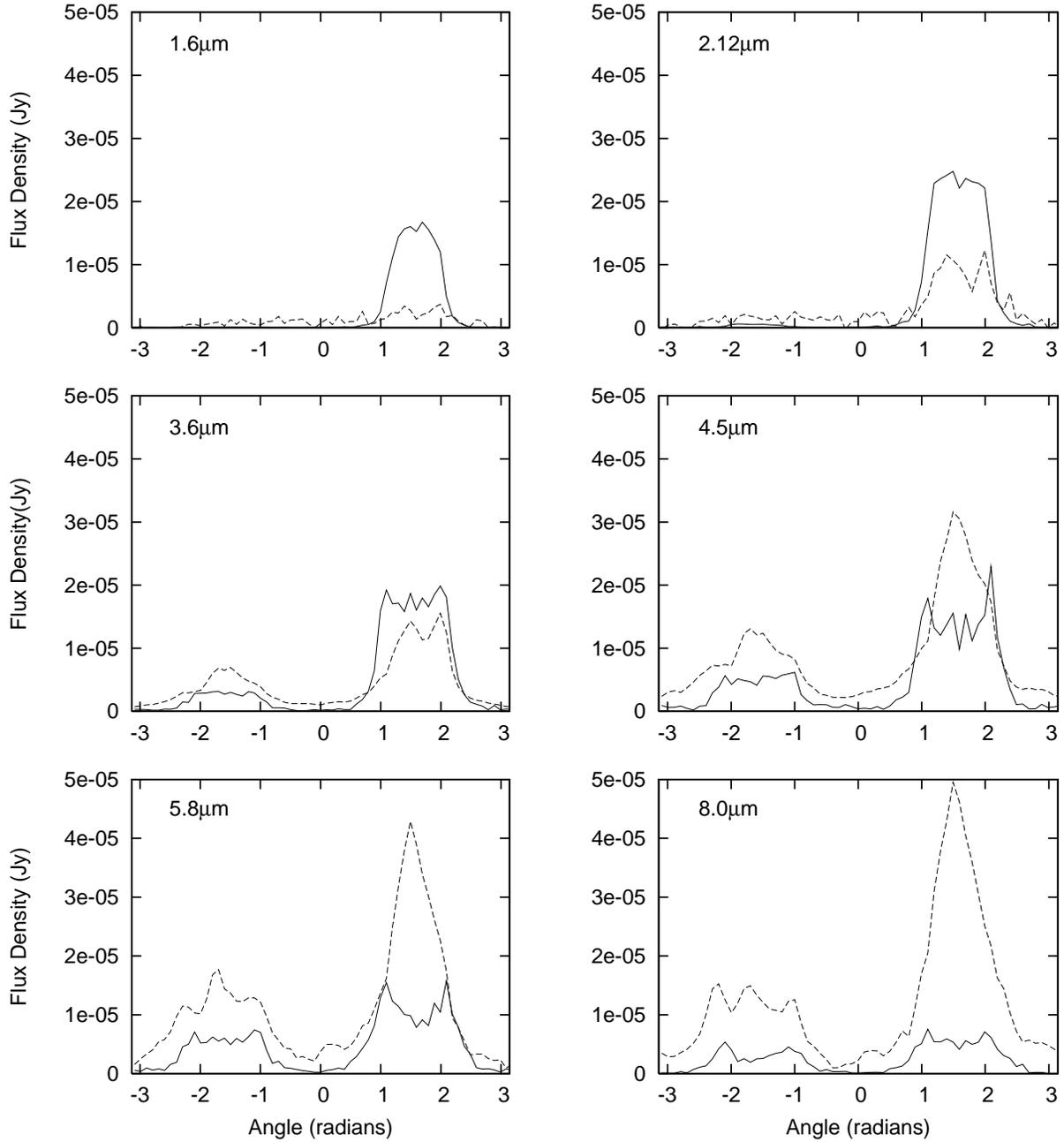}

\caption{Flux versus angle plots of IRS 2A with the best fit model. Models are solid lines, data are dashed lines. The entire angular range was used for fitting.}
\end{figure}

\begin{figure}
\figurenum{11}
\includegraphics[angle=-90, scale=.65]{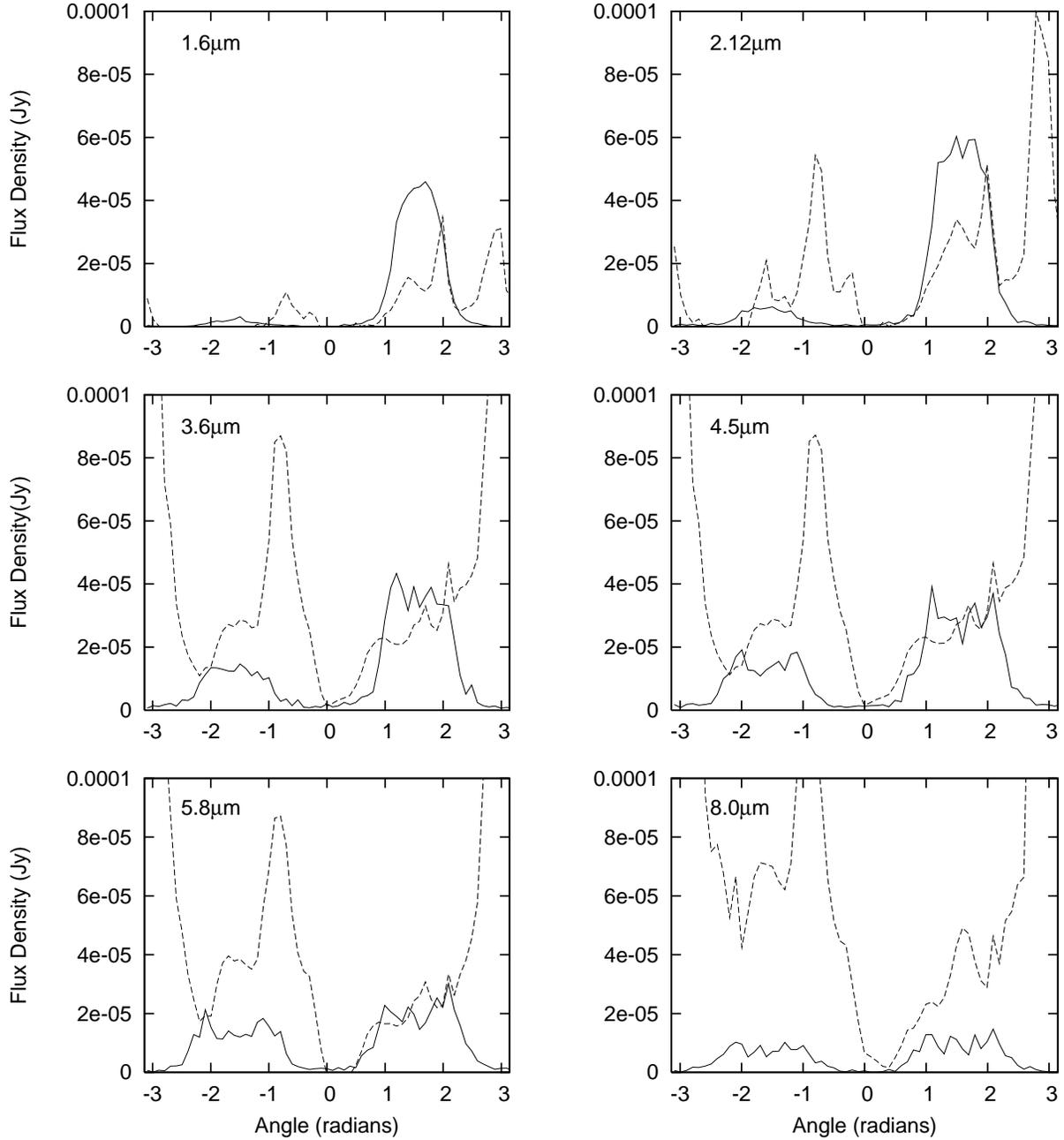}

\caption{Flux versus angle plots of IRS 3B with the best fit model. Models are solid lines, data are dashed lines.The center of the observed cavity is at $\sim1.57$ radians. The data from 0 to $\sim2.1$ radians were used for fitting. Confusion with IRS 3A and its outflow which contaminate $\sim2.1$ to 3.14 and 0 to -3.14.}
\end{figure}

\begin{figure}
\figurenum{12}
\includegraphics[angle=-90, scale=.65]{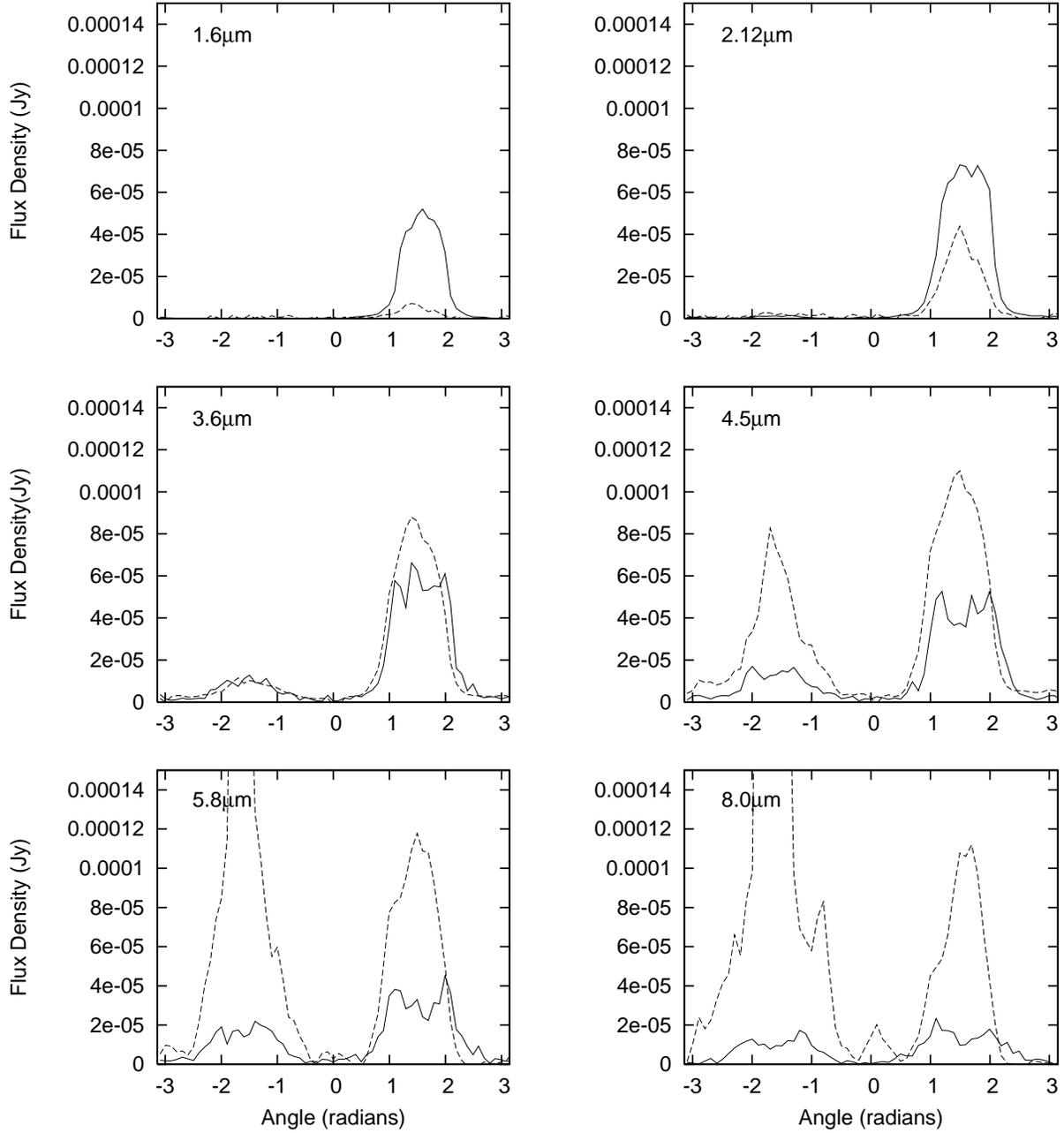}

\caption{Flux versus angle plots of L1448-mm A with the best fit model. Models are solid lines, data are dashed lines. Only the data from 0 to 3.14 radians were used for fitting because data from 0 to -3.14 are contaminated by the binary companion.}
\end{figure}

\begin{figure}
\figurenum{13}
\includegraphics[angle=-90, scale=.8]{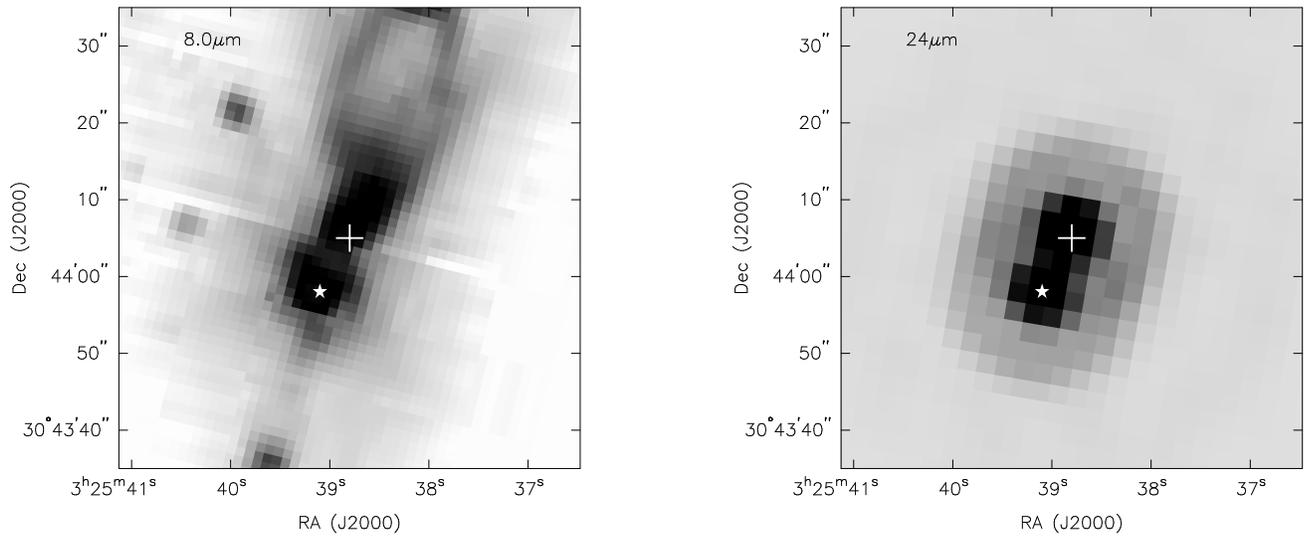}
\caption{IRAC channel 2 image of L1448-mm (left) compared to the MIPS 24$\mu$m image (right). 
L1448-mm A is marked with a cross and L1448-mm B is marked with a star. MIPS image courtesy of the \textit{cores2disks} Spitzer Legacy program.}
\end{figure}

\end{document}